\documentclass[twocolumn]{emulateapj}% change onecolumn to iop for fancy, iop to onecolumn for manuscript
\let\pwiflocal=\iffalse \let\pwifjournal=\iffalse
\usepackage{enumerate}
\usepackage{amsmath,amssymb}
\usepackage{bm}
\usepackage{color}
\usepackage[utf8]{inputenc}

\usepackage{amsmath,amssymb}
\usepackage[breaklinks,colorlinks,urlcolor=blue,citecolor=blue,linkcolor=blue]{hyperref}
\usepackage{epsfig}    
\usepackage{graphicx}    
\usepackage{lineno}
\usepackage{natbib}
\usepackage{bigints}
\usepackage[outdir=./]{epstopdf}

\usepackage[T1]{fontenc}
\pwifjournal\else
  \usepackage{microtype}
\fi

\pwifjournal\else
  \makeatletter
  \renewcommand\plotone[1]{%
    \centering \leavevmode \setlength{\plot@width}{0.95\linewidth}
    \includegraphics[width={\eps@scaling\plot@width}]{#1}%
  }%
  \makeatother
\fi

\makeatletter

\newcommand\@simpfx{http://simbad.u-strasbg.fr/simbad/sim-id?Ident=}

\newcommand\MakeObj[4][\@empty]{% [shortname]{ident}{url-escaped}{formalname}
  \pwifjournal%
    \expandafter\newcommand\csname pkgwobj@c@#2\endcsname[1]{\protect\object[#4]{##1}}%
  \else%
    \expandafter\newcommand\csname pkgwobj@c@#2\endcsname[1]{\href{\@simpfx #3}{##1}}%
  \fi%
  \expandafter\newcommand\csname pkgwobj@f#2\endcsname{#4}%
  \ifx\@empty#1%
    \expandafter\newcommand\csname pkgwobj@s#2\endcsname{#4}%
  \else%
    \expandafter\newcommand\csname pkgwobj@s#2\endcsname{#1}%
  \fi}%

\newcommand\MakeTrunc[2]{% {ident}{truncname}
  \expandafter\newcommand\csname pkgwobj@t#1\endcsname{#2}}%

\newcommand{\obj}[1]{%
  \expandafter\ifx\csname pkgwobj@c@#1\endcsname\relax%
    \textbf{[unknown object!]}%
  \else%
    \csname pkgwobj@c@#1\endcsname{\csname pkgwobj@s#1\endcsname}%
  \fi}
\newcommand{\objf}[1]{%
  \expandafter\ifx\csname pkgwobj@c@#1\endcsname\relax%
    \textbf{[unknown object!]}%
  \else%
    \csname pkgwobj@c@#1\endcsname{\csname pkgwobj@f#1\endcsname}%
  \fi}
\newcommand{\objt}[1]{%
  \expandafter\ifx\csname pkgwobj@c@#1\endcsname\relax%
    \textbf{[unknown object!]}%
  \else%
    \csname pkgwobj@c@#1\endcsname{\csname pkgwobj@t#1\endcsname}%
  \fi}

\makeatother

\pwifjournal\else
  \usepackage{etoolbox}
  \makeatletter
  \patchcmd{\NAT@citex}
    {\@citea\NAT@hyper@{%
       \NAT@nmfmt{\NAT@nm}%
       \hyper@natlinkbreak{\NAT@aysep\NAT@spacechar}{\@citeb\@extra@b@citeb}%
       \NAT@date}}
    {\@citea\NAT@nmfmt{\NAT@nm}%
     \NAT@aysep\NAT@spacechar\NAT@hyper@{\NAT@date}}{}{}
  \patchcmd{\NAT@citex}
    {\@citea\NAT@hyper@{%
       \NAT@nmfmt{\NAT@nm}%
       \hyper@natlinkbreak{\NAT@spacechar\NAT@@open\if*#1*\else#1\NAT@spacechar\fi}%
         {\@citeb\@extra@b@citeb}%
       \NAT@date}}
    {\@citea\NAT@nmfmt{\NAT@nm}%
     \NAT@spacechar\NAT@@open\if*#1*\else#1\NAT@spacechar\fi\NAT@hyper@{\NAT@date}}
    {}{}
  \makeatother
\fi

\newcommand{\vM}{\mathsf{M}}

\newcommand{\teff}{T_\textrm{eff}}
\newcommand{\teffa}{T_\textrm{hot}}
\newcommand{\teffb}{T_\textrm{cool}}
\newcommand{\logg}{\log g}
\newcommand{\Z}{[{\rm Fe}/{\rm H}]}

\newcommand{\vsini}{v \sin i}

\newcommand{\flam}{f_\lambda}
\newcommand{\vt}{ {\bm \theta}}

\newcommand{\PHOENIX}{{\sc Phoenix}}

\newcommand{\um}{$\mu$m}

\newcommand{\iancze}{{\sc C15}}

\providecommand{\adsurl}[1]{\href{#1}{ADS}}
\newcommand{\name}{LkCa 4 }
\newcommand{\project}[1]{\textsl{#1}}

\slugcomment{Accepted to ApJ}

\shorttitle{Starspots on LkCa 4}

\shortauthors{Gully-Santiago et al.}

\begin{document}
 
\title{Placing the spotted T Tauri star LkCa 4 on an HR diagram}

\author{Michael A. Gully-Santiago,\altaffilmark{1} Gregory J. Herczeg,\altaffilmark{1} Ian Czekala,\altaffilmark{2} Garrett Somers,\altaffilmark{3,18} Konstantin Grankin,\altaffilmark{4} Kevin R. Covey,\altaffilmark{5} J.F. Donati,\altaffilmark{6,7} Silvia H. P. Alencar,\altaffilmark{8} Gaitee A.J. Hussain,\altaffilmark{9,10} Benjamin J. Shappee,\altaffilmark{11,12} Gregory N. Mace,\altaffilmark{13} Jae-Joon Lee,\altaffilmark{14} T.~W.-S.~Holoien,\altaffilmark{15,16} Jessy Jose,\altaffilmark{1} Chun-Fan Liu\altaffilmark{17,1}}

\altaffiltext{1}{Kavli Institute for Astronomy and Astrophysics, Peking University, Yi He Yuan Lu 5, Haidian Qu, 100871 Beijing, People's Republic of China}
\altaffiltext{2}{KIPAC, Physics and Astronomy, Stanford University, Stanford, CA}
\altaffiltext{3}{Department of Physics and Astronomy, Vanderbilt University, 6301 Stevenson Center, Nashville, TN, 37235, USA}
\altaffiltext{4}{Crimean Astrophysical Observatory, Nauchny, Crimea 298409, Russia}
\altaffiltext{5}{Department of Physics \& Astronomy, Western Washington University, Bellingham WA 98225-9164, USA}
\altaffiltext{6}{CNRS, IRAP / UMR 5277, Toulouse, 14 avenue E. Belin, F- 31400 France}
\altaffiltext{7}{Université de Toulouse, UPS-OMP, IRAP, 14 avenue E. Belin, Toulouse, F–31400 France}
\altaffiltext{8}{Departamento de F\`{\i}sica -- ICEx -- UFMG, Av. Ant\^onio Carlos, 6627, 30270-901 Belo Horizonte, MG, Brazil}
\altaffiltext{9}{European Southern Observatory, Karl-Schwarzschild-Str. 2, 85748 Garching bei M\"unchen, Germany}
\altaffiltext{10}{Institut de Recherche en Astrophysique et Plan\'etologie, Universit\'e de Toulouse, UPS-OMP, F-31400 Toulouse, France}
\altaffiltext{11}{Carnegie Observatories, 813 Santa Barbara Street,
  Pasadena, CA 91101, USA}
\altaffiltext{12}{Hubble, Carnegie-Princeton Fellow}
\altaffiltext{13}{Department of Astronomy, The University of Texas at Austin, 2515 Speedway St, Austin, TX, USA}
\altaffiltext{14}{Korea Astronomy and Space Science Institute, 776 Daedeok-daero, Yuseong-gu, Daejeon 305-348, Korea.}
\altaffiltext{15}{Department of Astronomy, The Ohio State University, 140 West 18th Avenue, Columbus, OH 43210, USA}
\altaffiltext{16}{Center for Cosmology and AstroParticle Physics (CCAPP), The Ohio State University, 191 W. Woodruff Ave., Columbus, OH 43210, USA}
\altaffiltext{17}{Institute of Astronomy and Astrophysics, Academia Sinica (ASIAA), Taipei 10617, Taiwan}
\altaffiltext{18}{VIDA Postdoctoral Fellow}

\begin{abstract}

Ages and masses of young stars are often estimated by comparing their luminosities and effective temperatures to pre-main sequence stellar evolution tracks, but magnetic fields and starspots complicate both the observations and evolution. To understand their influence, we study the heavily-spotted weak-lined \emph{T-Tauri} star LkCa 4 by searching for spectral signatures of radiation originating from the starspot or starspot groups.  We introduce a new methodology for constraining both the starspot filling factor and the spot temperature by fitting two-temperature stellar atmosphere models constructed from \PHOENIX\ synthetic spectra to a high-resolution near-IR IGRINS spectrum.  Clearly discernable spectral features arise from both a hot photospheric component $\teffa\sim4100$ K and to a cool component $\teffb\sim2700-3000$ K, which covers $\sim80\%$ of the visible surface.   This mix of hot and cool emission is supported by analyses of the spectral energy distribution, rotational modulation of colors and of TiO band strengths, and features in low-resolution optical/near-IR spectroscopy.   Although the revised effective temperature and luminosity make LkCa 4 appear much younger and lower mass than previous estimates from unspotted stellar evolution models, appropriate estimates will require the production and adoption of spotted evolutionary models.  Biases from starspots likely afflict most fully convective young stars and contribute to uncertainties in ages and age spreads of open clusters.  In some spectral regions starspots act as a featureless ``veiling'' continuum owing to high rotational broadening and heavy line-blanketing in cool star spectra.  Some evidence is also found for an anti-correlation between the velocities of the warm and cool components.
\end{abstract}

\keywords{stars: fundamental parameters --- stars: individual (LkCa 4) ---  stars: low-mass -- stars: statistics}

\maketitle
\section{Introduction}\label{sec:intro}
The causes of age uncertainties and of large luminosity spreads in young clusters are controversial \citep[see reviews by][]{preibisch12,soderblom14}.  Absolute ages are usually inferred from comparisons between observed stellar properties and pre-main sequence evolutionary models, which are sensitive to stellar birthlines and the input physics that controls the contraction and internal heating.  Within clusters, any real differences in ages between stars may be masked by observational uncertainties in temperature and luminosity measurements \citep[e.g.][]{hartmann01,slesnick08}.

Magnetic activity is a likely source of significant uncertainty in both the models and observations of young low-mass stars. Convection at these young ages generates strong magnetic fields, as measured by Zeeman broadening and polarimetry \citep[e.g.][]{johnskrull07,donati09} and as seen in starspots \citep[e.g.][]{stauffer03,grankin08}.  Evolutionary models are just now starting to implement prescriptions for how spots and magnetic fields affect the interior structure \citep{macdonald13,jackson14a,somers15,feiden16}.  Stellar evolution models that include these effects can make a coeval 10 Myr population exhibit apparent age spreads of 3$-$10 Myr.  Mass estimates may also be biased to lower masses, as measured for some eclipsing binaries \citep[e.g.][]{stassun14,rizzuto16}.   Observationally, starspots may also be responsible for biases in stellar effective temperatures derived by different methods.  For example, the effective temperatures for 3493 young stars measured using the APOGEE spectrograph \citep[$1.5-1.70 \;\mu$m at $R=22,500$][]{wilson10} are offset from and usually cooler than prior measurements made at optical wavelengths \citep{cottaar14}.  Similarly, near-IR spectral types of young stars have sometimes been evaluated as later than optical spectral types \citep{bouvier92,vacca11}.

Rotational modulation from starspots is understood as the cause of cyclic variations in weak-lined T Tauri stars (WTTSs) and some classical T Tauri stars \citep[e.g.][]{vrba86,herbst94}.  Spots have recently been observed with exceptional photometric precision from monitoring surveys targeting exoplanet transits \citep[\emph{e.g}][]{harrison11,davenport15}.  The single band light curves from such surveys cannot separate the relative areas and temperatures of the emitting regions, and therefore cannot yield estimates for the filling factor of spots without making assumptions about the relative temperature contrast. Contemporaneous or near-contemporaneous panchromatic photometric monitoring \citep{herbst94,petrov94,bouvier95,grankin07,cody14} can break the degeneracy between starspot filling factor and starspot temperature, but these measurements are still limited to detecting large longitudinally asymmetric starspots.  Transiting planets or eclipsing binaries passing in front of starspots \citep{desert11} can break geometric degeneracies, providing estimates for the characteristic sizes and lifetimes of starspots or starspot groups, but the number of such transiting systems and demands of high photometric precision limit wider application.  Zeeman Doppler Imaging \citep[ZDI]{donati14} produces the latitudinal and longitudinal distribution of starspots in reconstructed brightness maps based on models for spectro-polarimetric line shapes but would be unable to identify uniform distributions of small spots.  All of these techniques would tend to underestimate the fractional coverage of starspots distributed isotropically on the surface, stars viewed pole-on, or stars with longitudinally symmetric bands of starspots.

Spectroscopic strategies have the power to measure both the starspot areal coverage and temperature in any geometric configuration.  For WTTSs, SED modeling indicated the presence of cool spots \citep{wolk96}, while spot temperatures have been estimated from rotational modulation of optical colors \citep[e.g.][]{grankin98,venuti15,koen16}.  The TiO bands in optical spectra have been used to measure spots on WTTSs \citep{petrov94}, stars with ages of 125 Myr in the Pleiades \citep{fang2016}, and evolved stars \citep[e.g.][]{neff95,oneal01,oneal04}.  Recent spectroscopic detections of spots on young stars have focused on TW Hya, an active accretor, and DQ Tau, a close binary in which both components are accreting \citep{debes13,bary14}.  

The star LkCa 4, a WTTS member of the Taurus Molecular Cloud \citep{herbig86,strom89a,downes88,strom89b} is an ideal exemplar for a spotted pre-MS star because it does not have any mid-IR or mm excess \citep[\emph{e.g.}][]{andrews05,furlan06,buckle15} and is not actively accreting \citep[\emph{e.g.}][]{edwards06,cauley12}.  LkCa 4 has no evidence for a close companion from either direct imaging searches \citep{karr10,kraus11,daemgen15}\footnote{The status for wide companions is less clear; see \citet{stauffer91,itoh08,kraus09,kraus11,herczeg14}} or from RV searches \citep{nguyen12} and high-precision Zeeman Doppler Imaging \citep{donati14}.  This single star demonstrates a large amplitude of sinusoidal photometric variability indicative of large spots \citep{grankin08,xiao12}.  The variability amplitude cannot arise from an eclipsing binary since the large RV variations would have been seen in spectroscopic monitoring.  All of these observations indicate that the spectrum of \name should be devoid of complicating factors like near-IR excess veiling, accretion excess, or close binarity.    

LkCa 4 has recently been examined with ZDI \citep{donati14}, revealing a complex distribution of cool and warm spots in brightness map reconstructions.  Dark polar spots extend to about $30^\circ$ from the pole, with about 5 appendages reaching down to about $60^\circ$ from the pole.  There is also evidence for a hot spot in the reconstructed ZDI map.  These spot maps demonstrate that multiple temperature components contribute to the broadband emission from the star, which may explain the large differences between the 4100 K effective temperature measurement obtained from high-resolution optical spectra \citep{donati14} compared with 3600 K obtained from an analysis dominated by TiO bands \citep{herczeg14}.

In this work, we measure the starspot properties of LkCa 4 with four complementary techniques.\footnote{Some data, Python code, Starfish configuration files, and other documents related to this paper are available on the project's GitHub repository at \url{https://github.com/BrownDwarf/welter}}  
Section \ref{sec:obs} describes the full spectroscopic and photometric dataset and 
quantifies the optical variability of LkCa 4 over the last 31 years based on all available photometric monitoring; a collection of spectral observations at various phases of variability is introduced.  Section 3 and Appendix \ref{methods-details} describe a spectral inference technique aimed at generalizing line depth ratio analysis, which is applied to a high resolution $H$ and $K$ near-IR spectrum of LkCa 4 in Section \ref{sec:two_tempIGRINS}.  Section \ref{sec:GJHsection4} shows results from SED fitting, polychromatic photometric monitoring, and optical TiO line-depth ratio fitting.  All lines of evidence are ultimately combined to build a consistent picture of the spectral and temporal evolution of LkCa 4.  Section 5 describes how the two-temperature fit affects the placement of LkCa 4 on the HR diagram and what can be understood about pre-main sequence stellar evolution from this exemplar.  Finally, Section 6 describes how our results should be interpreted, the implications for analyses of spots in high-resolution spectra, and methodological limitations.

\section{Observations}\label{sec:obs} 

\subsection{IGRINS Spectroscopy}\label{sec:igrins} 
We acquired observations of LkCa 4 with the Immersion Grating Infrared Spectrograph, IGRINS \citep{park14} on the Harlan J. Smith Telescope at McDonald Observatory on 2015-11-18 $09^h$ UTC.  IGRINS is a high resolution near-infrared echelle spectrograph providing simultaneous $R\simeq45,000$ spectra over 1.48-2.48 \um.  The spectrograph has two arms with 28 orders in $H-$band and 25 orders in $K-$band.  The data were reduced with the Pipeline Package \citep{jaejoonlee15}.  Telluric correction was performed by dividing the spectrum of \name by a spectrum of HR 1237, an A0V star observed immediately before \name, both at airmass 1.1.  The broad hydrogen lines in the A0V star produced excess residuals in the spectrum of LkCa 4.  No effort was made to mitigate these flux excess residuals, which affect several spectral orders.  These orders were ultimately excluded from further analyses.

\subsection{ESPaDOnS Spectroscopy}
We used ESPaDOnS on CFHT to obtain twelve high-resolution optical spectra of \name from 8-21 Jan.~2014 as part of the MaTYSSE Large Program.  These spectra cover 3900$-$10000 \AA\ at $R\sim68,000$ and were obtained in spectropolarimetry mode.  The Zeeman Doppler Imaging obtained from these observations were analyzed by \citet{donati14}.  In this paper, we concentrate on the intensity spectrum.  To calculate TiO indices, the relative flux calibration was obtained from an approximate instrument sensitivity calculated from observations of 72 Tau.  The spectral shapes match the flux calibrated low-resolution spectra described in \S 2.3.

\subsection{Low-resolution optical and near-IR spectra}

A flux-calibrated low-resolution optical spectrum of LkCa 4 was obtained using Palomar/DBSP \citep[DBSP,][]{oke82} on 30 Dec. 2008 at 06:30 UT.  These spectra cover 3200--8700 \AA\ at a spectral resolution $R\sim 500$.  The data reduction and a spectral analysis are described by \citet{herczeg14} within the context of a larger survey.

Near-infrared spectra of LkCa 4 were obtained at ~01:30 UT on Dec. 30, 2008, a few hours earlier than the DBSP spectra, using the TripleSpec spectrograph on the Astrophysical Research Consortium (ARC) 3.5 meter telescope at Apache Point Observatory in Sunspot, New Mexico. LkCa 4 was observed for a total integration time of 90 s using an ABBA dither sequence along the slit to remove sky emission by differencing sequential exposures.  With the 1.1$\arcsec$ slit used for these observations, TripleSpec provides nearly contiguous coverage from 0.95 to 2.46 $\mu$m at a resolution of  R $\sim$3200 \citep{wilson04}. Spectra were reduced with a variant of the SpeXTool pipeline, originally developed by \citet{cushing04} and modified for use with TripleSpec data. Spectra were differenced, flattened, extracted, and wavelength calibrated prior to telluric correction and flux calibration; the latter two corrections were performed using the XTELLCOR IDL package \citep{vacca03} and a spectrum of HD 25175, a nearby A0V star observed directly after LkCa 4 at a similar airmass.

\subsection{Photometric monitoring}

\begin{figure*}
 \centering
 \includegraphics[width=0.98\textwidth]{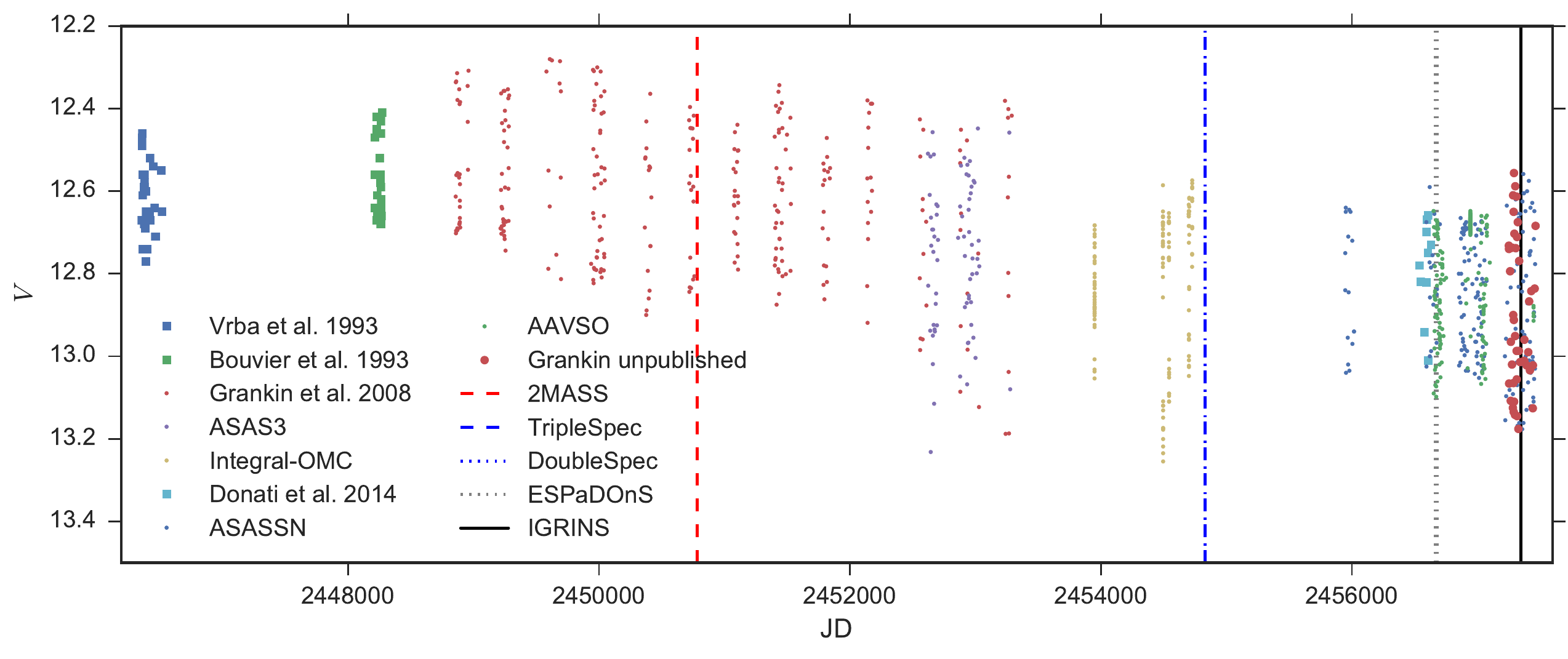}
 \caption{Overview of \name $V-$band photometric monitoring from 1986$-$2016.  The vertical lines denote the observing epochs of 2MASS, IGRINS, ESPaDOnS, DBSP, and TripleSpec.  The near contemporaneous DBSP and TripleSpec epochs lay on top of each other on this scale, as do the 12 ESPaDOnS epochs.  The abscissa range is equal to the current lifespan of the first author of this paper.}
 \label{fig:PhotTime}
\end{figure*}

\begin{figure*}
 \centering
 \includegraphics[width=0.95\textwidth]{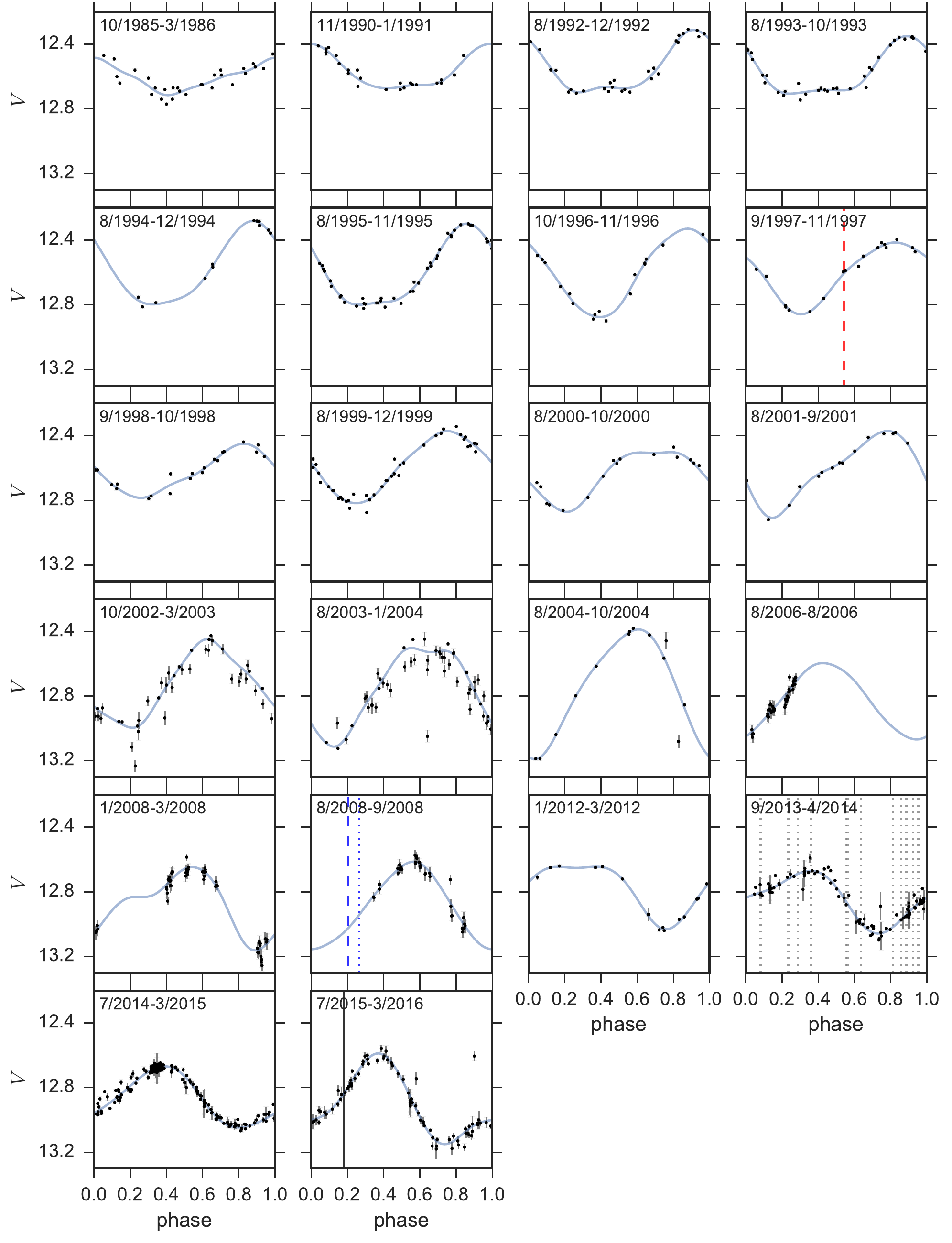}
 \caption{Phase-folded lightcurves constructed assuming $P=3.375$ days for all 22 observing seasons.  The blue solid lines show a ``multiterm'' regularized periodic fit, that is, keeping the first $M_{\rm max}=4$ Fourier components \citep{vanderplas15a}.  The vertical lines show the epochs at which spectra or ancillary photometry were obtained, with the same line styles and colors as Figure \ref{fig:PhotTime}.  LkCa 4 shows secular changess in its light curve morphology.  The IGRINS spectrum was acquired near the median flux level of the season.}
 \label{fig:PhotPhase}
\end{figure*}

Photometry in $B$, $V$, and/or $R$-bands in 1216 distinct epochs was assembled from published and unpublished data sources spanning 31 years in 22 observing seasons.  Published data targeting \name is composed of 29 $BVR$ visits from 1985$-$1986 \citep{vrba93}, 26 $BVR$ epochs from 1990$-$1991 \citep{bouvier93}, 284 visits in $V$-band (278 with $B$ and 268 with $R$) from 1992 August - 2004 October \citep{grankin08}, and 10 $BVR$ visits from 2013 \citep{donati14}.  ASAS3 \citep{pojmanski04} acquired 63 $V-$band measurements from 2002$-$2004.  Integral-OMC obtained 138 $V-$band measurements from 2006$-$2008 \citep{garzon12}.  The AAVSO archive \citep{kafka16} includes 385 $V$, 23 $B$ and 10 $R$ measurements from 2013$-$2016.  Unpublished data from the ASAS-SN survey \citep{shappee14} were obtained from 2012 January -- 2016 March in 186 visits.  Finally, recent photometry from the ongoing Crimean Astrophysical Observatory (CrAO) ROTOR project \citep{grankin08} is presented here for the first time.  The CrAO data includes 43 $BVR$ visits from August 2015 - March 2016. Figure \ref{fig:PhotTime} displays all the $V-$band photometry over the interval 1985-2016.  HJD, BJD, and JD are simply referred to as JD, which is accurate to the precision of available data.

Figure \ref{fig:PhotPhase} shows all available $V-$band data grouped by the 22 observing seasons and phase-folded by the period $P=3.375$ days obtained from multiterm Lomb-Scargle periodograms \citep{ivezic14}.  The general appearance of the phase-folded lightcurves does not change with perturbations to the period on the scale of 0.003 days.  The estimated $BVR$ magnitudes at the time of the spectral observations are determined from regularized multiterm fits \citep[\emph{i.e.} Fourier series truncated to the first $\sim 4$ components]{vanderplas15a} shown as the solid blue lines in Figure \ref{fig:PhotPhase}.  Table \ref{tbl_estimated_V} lists the estimated $BVR$ photometry and the observing epoch for data from IGRINS, ESPaDOnS, 2MASS \citep{skrutskie06}, DBSP, and TripleSpec.

\begin{deluxetable}{rrccc}

\tabcolsep=0.11cm
\tablecaption{Estimated $BVR$ magnitudes\label{tbl_estimated_V}}
\tablewidth{0pt}
\tablehead{
\colhead{JD $-$ 2450000} &
\colhead{Instrument} &
\colhead{$B$} &
\colhead{$V$} &
\colhead{$R$}
}
\startdata
  781.7106 &       2MASS & 14.02 & 12.61 & 11.25 \\
 4830.5637 &  TripleSpec &   $\cdots$ & 13.02 &   $\cdots$ \\
 4830.7719 &  DoubleSpec &   $\cdots$ & 12.94 &   $\cdots$ \\
 6665.7204 &    ESPaDOnS & 14.36 & 12.87 & 11.71 \\
 6666.8505 &    ESPaDOnS & 14.15 & 12.69 & 11.65 \\
 6667.7727 &    ESPaDOnS & 14.31 & 12.87 & 11.63 \\
 6668.8699 &    ESPaDOnS & 14.44 & 12.94 & 11.75 \\
 6672.8995 &    ESPaDOnS & 14.25 & 12.80 & 11.56 \\
 6673.8408 &    ESPaDOnS & 14.12 & 12.67 & 11.62 \\
 6674.7746 &    ESPaDOnS & 14.46 & 12.99 & 11.85 \\
 6675.7396 &    ESPaDOnS & 14.40 & 12.90 & 11.73 \\
 6676.7954 &    ESPaDOnS & 14.18 & 12.72 & 11.65 \\
 6677.8699 &    ESPaDOnS & 14.29 & 12.86 & 11.61 \\
 6678.7419 &    ESPaDOnS & 14.53 & 13.02 & 11.80 \\
 6678.8950 &    ESPaDOnS & 14.47 & 12.97 & 11.75 \\
 7344.8610 &      IGRINS & 14.24 & 12.84 & 11.38 \\
\enddata
%\tablecomments{}
\end{deluxetable}

\section{FITS TO HIGH RESOLUTION SPECTRA}\label{sec:Starfish}

\begin{figure}
 \centering
 \includegraphics[width=0.48\textwidth]{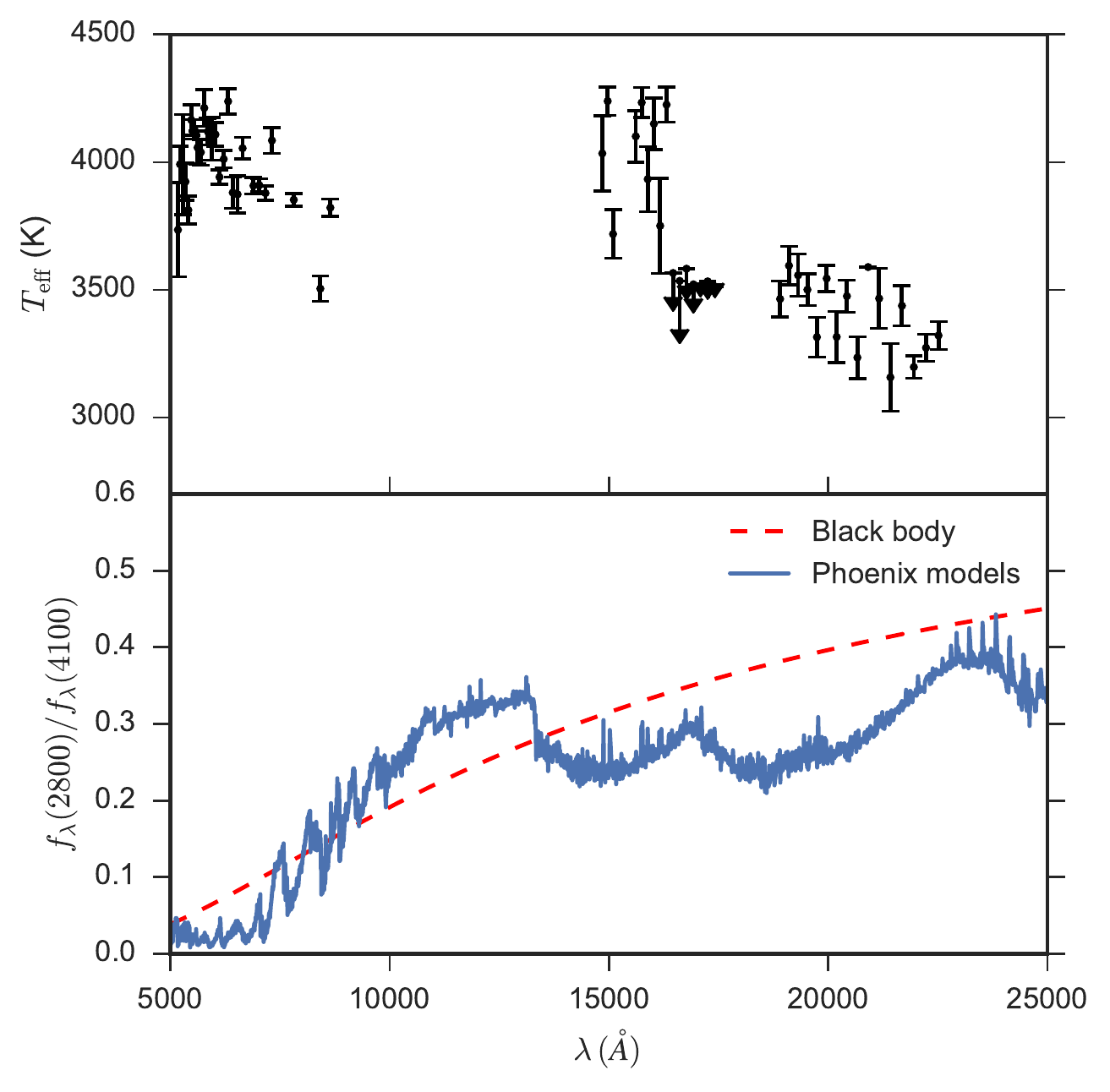}
 \caption{Top panel:  the best-fit temperature measured from our full spectrum fitting to each of 58 spectral orders in the optical and infrared portions of the spectrum,  assuming a single component photosphere.  The best-fit temperature measured in the $K-$band is about 800 K lower than that derived in optical.  The triangles in the $H$-band show upper limits with best fits near the edge of our grid.  Bottom panel:  The flux ratio between the cool and hot components increases to the near-IR, which explains the observed trend in the best-fit temperature.}
 \label{fig:SingleTeffvsOrder}
\end{figure}

Starspots possess a spectrum distinct from the ambient photosphere.  Approaches like ZDI and line-bisector analysis \citep[\emph{e.g.}][]{prato08, donati14} target the modulation of stellar photospheric lines as starspots enter and exit the observable stellar disk.  In contrast, the technique described in this Section measures the spectrum that emerges from both the hot and the cool surfaces of the photosphere.

Starspot line depth ratio analysis has traditionally been limited to isolated portions of spectrum that possess absorption lines attributable only to cool photospheric components and spectral lines arising from the warmer ambient photosphere \citep[\emph{e.g.}][]{neff95, oneal01}.  The apparent veiling of the lines and their respective temperature dependences can be combined to solve for the cool photosphere and ambient photosphere temperatures and relative areal coverages.  This line depth ratio method suffers from the need to identify portions of the spectrum that possess easily identifiable strong lines, and from the need to assemble large atlases of observed spotless spectral templates to which the spotted star spectrum can be compared.  In this section, we introduce a generalization of the line depth ratio analysis which employs pixel-by-pixel modeling of all echelle spectral orders.  By using all the spectral data, this strategy has the power to constrain starspot properties at relatively low filling factors and to identify weak lines that originate from the cool photosphere.

We took two approaches to characterizing the photospheric temperature from a spotted-star spectrum.  First, 
using the spectral fitting approach introduced in Section \ref{sec:methods}, we separately fit the optical (ESPaDOnS) and near-IR (IGRINS) spectra with distinct, single photospheric component models; these fits are described in Sections \ref{sec:ESP_starfish} and \ref{sec:IGR_starfish} respectively, and then compared in \ref{sec:whyNearIR}.
Second, we use a two temperature mixture model to perform a fit to the near-IR spectra in Section \ref{sec:two_tempIGRINS}.  The near-IR spectral range is preferred over the optical, since the cool photosphere emits most of its flux in the near-IR, enhancing the likelihood of direct detection of emission from the cool component.

\subsection{Methodology}\label{sec:methods} 

\citet[hereafter \iancze]{czekala15} developed a modular framework\footnote{The open source codebase and its full revision history is available at \url{https://github.com/iancze/Starfish}.  The experimental fork discussed in this paper is at \url{https://github.com/gully/Starfish}.  } to infer stellar properties from high-resolution spectra.  The \iancze\ technique forward models observed spectra with synthetic spectra from pre-computed model grids.  The intra-grid-point spectra are ``emulated'' in a process similar but superior to interpolation since it seamlessly quantifies the uncertainty attributable to the coarsely sampled stellar intrinsic parameters (see Appendix of \iancze).  The forward model includes calibration parameters, line spread functions, and a Gaussian process noise model to account for correlations in the residual spectrum.  We employed the \PHOENIX\ grid of pre-computed synthetic stellar spectra, which are provided at high spectral resolution over a wide wavelength range at 100 K intervals in stellar photospheric temperature\footnote{Single component \PHOENIX\ photospheres are indexed by the label ``effective temperature'', or $\teff$, which presupposes that the emergent spectrum is comprised of only a single photospheric component.  In this work we avoid labeling the photospheric components by $\teff$, and instead, refer to them as simply ``photospheric temperatures''.} in our range of interest \citep{husser13}.  The modular framework was altered to accommodate starspot measurements in two ways. First, the single photospheric component was updated to include two photospheric components.  Second, the MCMC sampling strategy was altered to accommodate the additional free parameters added by the two component model.  

The stellar photosphere is characterized as two photospheric components with a cool temperature $\teffb$ and hot temperature $\teffa$, with scalar solid angular coverages $\Omega_{\mathrm{cool}}$ and $\Omega_{\mathrm{hot}}$, respectively\footnote{For flux calibrated spectra, the total solid angle $\Omega$ can be constrained, but for typical echelle spectrographs only relative values of $\Omega_{\mathrm{cool}}$ and $\Omega_{\mathrm{hot}}$ can be inferred.}.  
The choice of nomenclature depends on the physical origin of the emission region, which is not directly observable without ancillary information.  For many applications, it could be reasonably assumed that the cool photospheric component corresponds to a sunspot or sunspot group, and the hot component corresponds to the ambient photosphere.  However, a cool ambient photosphere possessing localized hot spots---like plages on the sun---could yield the same or similar two-temperature composite spectrum.  In principle, the physical origin could be assessed with detailed comparison of average surface magnetic field strengths through Zeeman line broadening or other magnetic-field sensitive techniques.  For now, we index the two photospheric components in the most general way possible, simply \emph{cool} and \emph{hot}.

The two photospheric components share the same average intrinsic and extrinsic stellar parameters $\vsini, \logg, \Z, v_z$.  The composite mixture model for observed flux density is:
\begin{eqnarray} \label{eqn:mix_M}
S_{\mathrm{mix}} = \Omega_{\mathrm{hot}} B(\teffa)  + \Omega_{\mathrm{cool}} B(\teffb)
\end{eqnarray}
where $B(T)$ is the spectral radiance from model spectra.  The filling factor of the starspots is:
\begin{eqnarray} \label{eqn:fill_factor}
f_{\mathrm{cool}} = \frac{\Omega_{\mathrm{cool}}}{\Omega_{\mathrm{hot}} + \Omega_{\mathrm{cool}}}
\end{eqnarray}

The term $f_{\mathrm{cool}}$ represents an instantaneous, observational fill factor seen on one projected hemisphere, not $f_{\rm spot}$, the ``ratio of the spotted surface to the total surface areas'' \citep{somers15}.  In the limit of homogeneously distributed starspots and assuming the cool photosphere arises from starspots, $f_{\mathrm{cool}}$ tends to $f_{\rm spot}$, but in general $f_{\rm spot}$ is not a direct observable since some circumpolar regions of inclined stars will always face away from Earth; $f_{\mathrm{cool}}$ can vary cyclically through rotational modulation, whereas $f_{\rm spot}$ changes secularly through starspot evolution that is not yet understood.

The two component mixture model includes two more free parameters than the standard \iancze\ model, namely $\teffb$ and $f_{\mathrm{cool}}$.  The addition of these two parameters makes the MCMC sampling much more correlated than it was for single component models.  This challenge motivated a second technical change to the spectral inference framework, switching from Gibbs sampling (\emph{c.f.} \iancze) to ensemble sampling using \texttt{emcee} \citep{foreman13}.  The practical effect of this switch is that all 14 stellar and nuisance parameters are fit simultaneously in a single spectral order, making stellar parameter estimates $\vt_{o}$ unique for each spectral order $o$, whereas the \iancze\ strategy had the power to provide a single set of stellar parameters $\vt$ that was based on all $N_{\rm ord}$ spectral orders.

Interpreting the two-temperature mixture model as arising from the same photosphere requires single stars, or at least double stars with extremely large (>100) luminosity ratios or wide separations.  Unresolved single-lined or double-lined binaries could mimic a signal that would be misattributed to starspots.  As pointed out earlier, LkCa 4 has no detected companion from direct imaging, and its RV variations are consistent with starspot modulation, so any binary companion would have to have an exceptionally large luminosity (or mass) ratio, rendering it undetectable with our spectral inference methodology with the finite signal to noise available in the IGRINS data.  The absence of a detectable disk around LkCa 4 means that the spectral modeling requires no further parameters like veiling or accretion.

Further details of the spectral inference methodology are described in Appendix \ref{methods-details}.

\begin{figure*}
 \centering
 \includegraphics[width=0.45\textwidth]{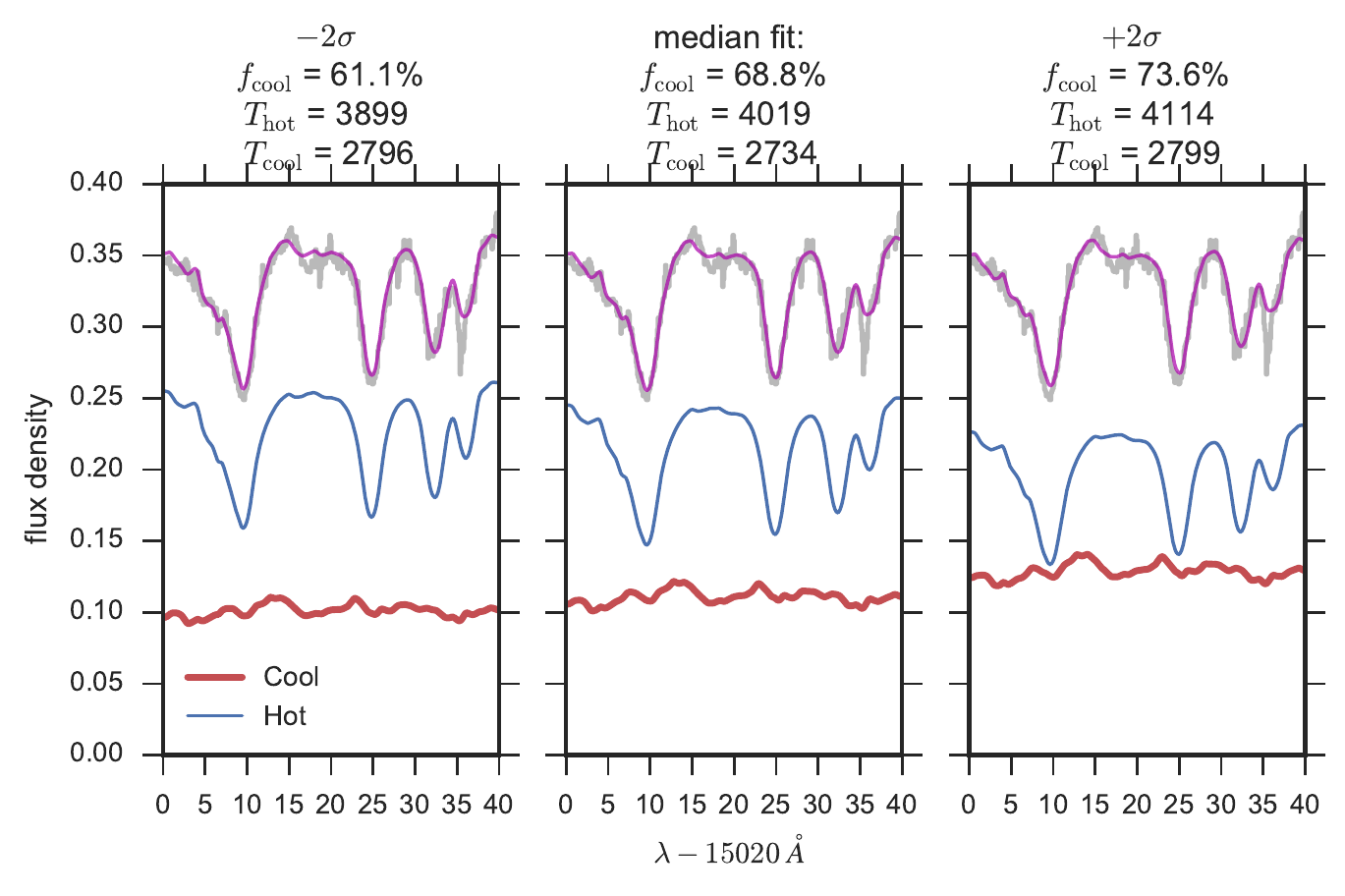} 
 \includegraphics[width=0.45\textwidth]{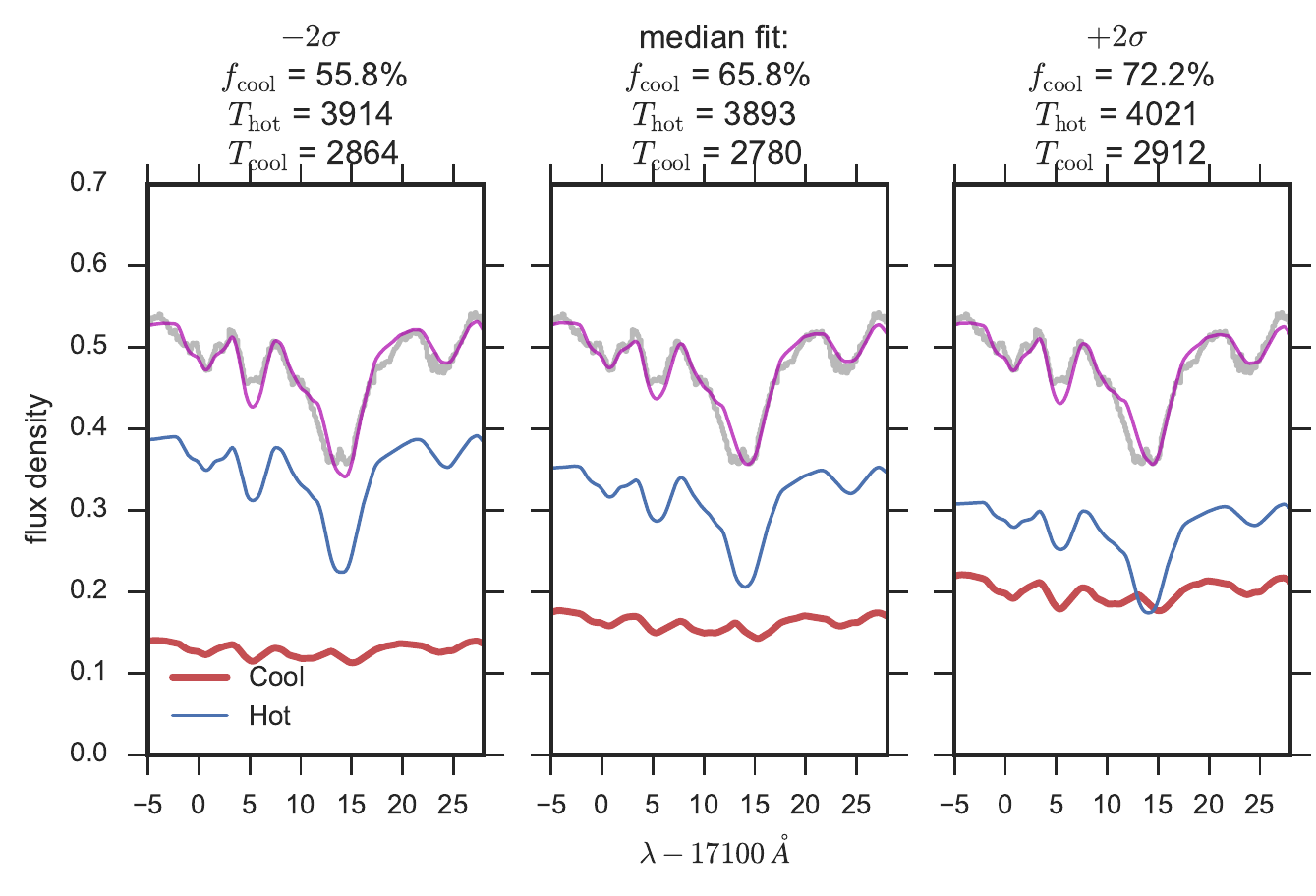} 
 \includegraphics[width=0.45\textwidth]{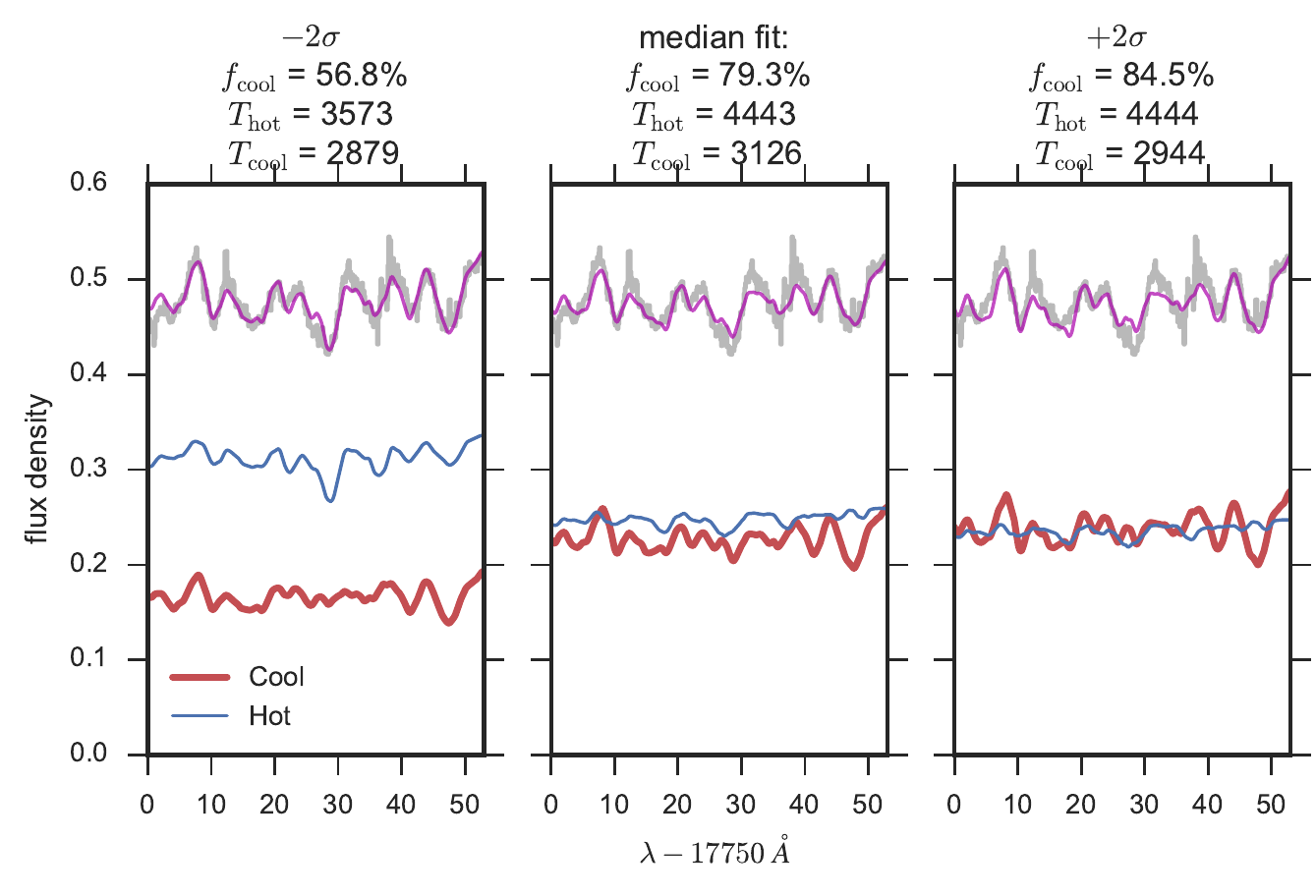} 
 \includegraphics[width=0.45\textwidth]{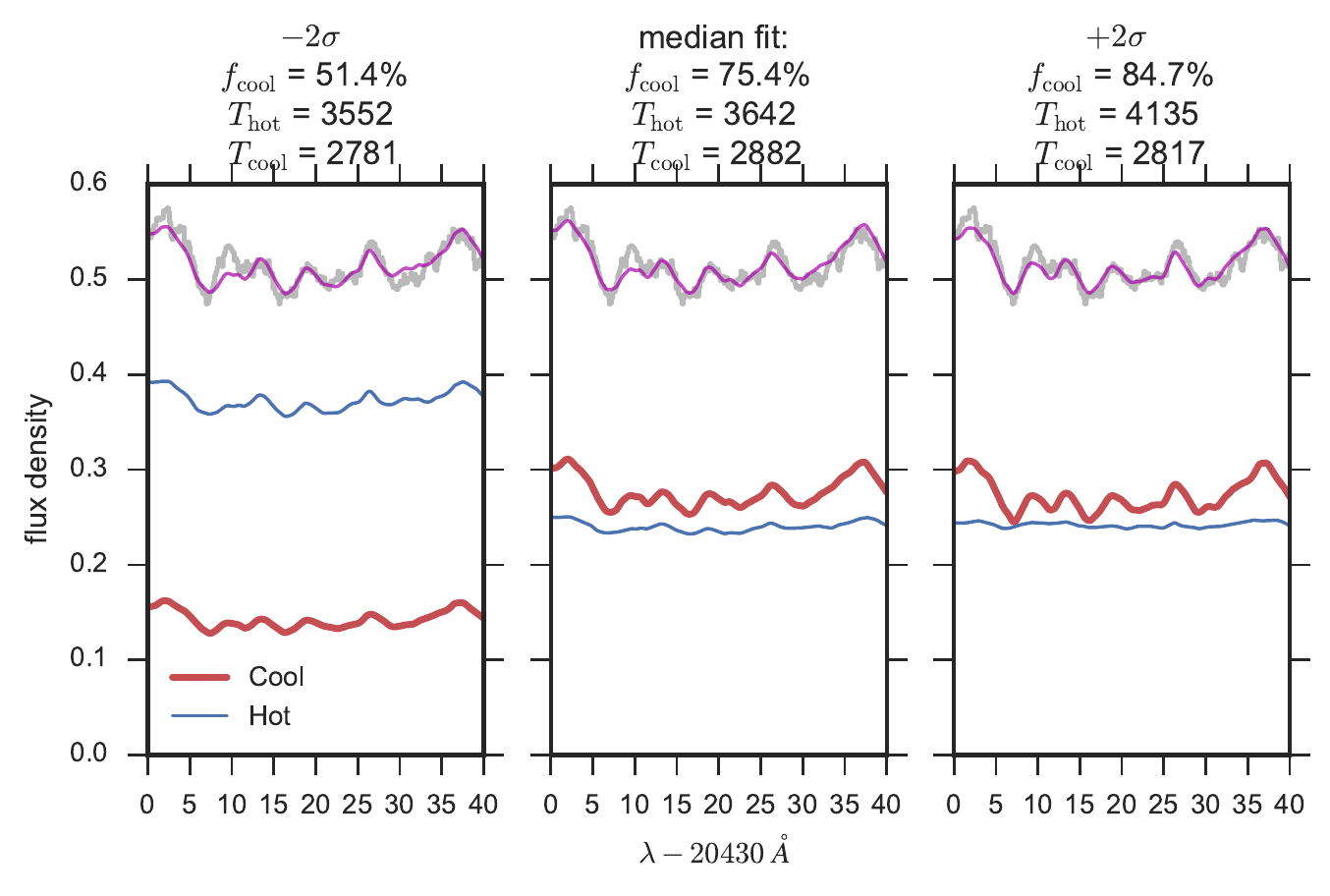} 
 \includegraphics[width=0.45\textwidth]{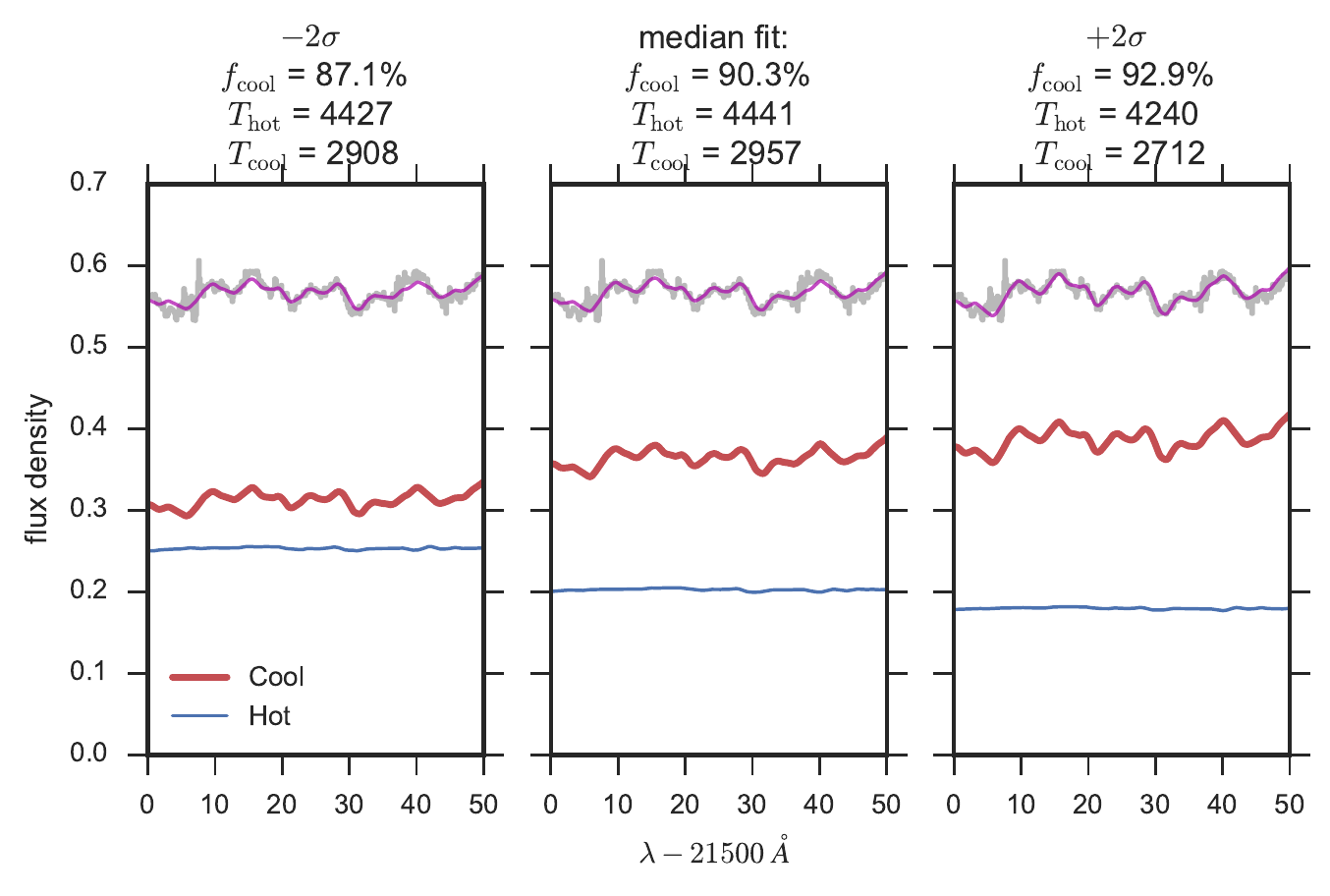} 
 \includegraphics[width=0.45\textwidth]{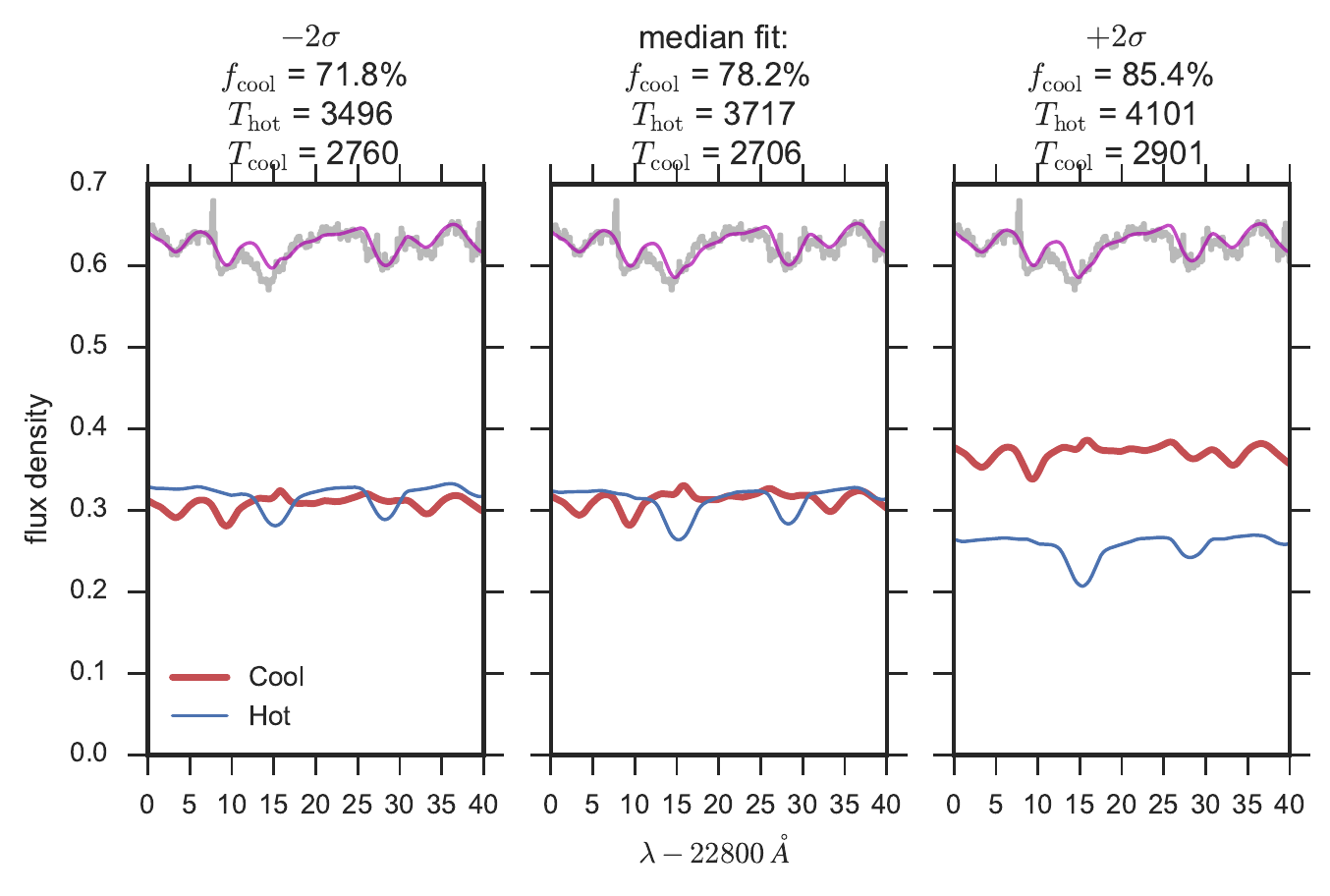} 
 \caption{Examples of spectral features in the observed IGRINS spectrum (light gray line).  The composite spectrum model (purple thin line) is consistent with the observed spectrum for a range of fill factors, with examples of the median fill factor (middle panel of triptych) and $\pm2\sigma$ fill factors demarcated on the spectral postage stamps.  The upper right triptych shows a Zeeman-sensitive Mg I line that is modeled with no attention to magnetic field, which may bias estimates of $\teffa$ and/or $f_{\mathrm{cool}}$ for individual spectral orders.}
 \label{fig:specPostageStamp}
\end{figure*}

\subsection{Single temperature fitting to the ESPaDOnS spectrum}\label{sec:ESP_starfish}

We performed spectral fitting on an ESPaDOnS spectrum acquired on 2014 January 11 (JD 2456668.9).  The spectrum was split into subsets of $N=26$ chunks, corresponding approximately to spectral order boundaries.  Fitting was performed separately on each spectral chunk.  The spectral emulator was trained on stellar parameters in the range $\logg \in [3.5, 4.0]$, $\Z \in [-0.5, 0.5]$, and photospheric temperatures $ T \in [3500, 4200]$ K.

Standard MCMC convergence criteria were assessed.  Typical spectral chunks show photospheric temperature point estimates in the range $4000\pm130$ K.  Figure \ref{fig:SingleTeffvsOrder} displays the point estimates with 5$^{th}$ and $95^{th}$ percentile error bars placed at the central wavelength of each spectral chunk.  The spectral chunk from 8473$-$8707 \AA\ is an outlier, showing an estimated photospheric temperature of $\sim3500$ K.

\subsection{Single temperature fitting to the IGRINS spectrum}\label{sec:IGR_starfish}

We selected a subset of 32 of the 54 available IGRINS spectral orders\footnote{The prohibitive computational cost limited us from running all the spectral orders.} with low telluric absorption artifacts.  We performed unique spectral fitting on each order, with free stellar parameters $\vt = (\teff, \logg, \Z)$.  The same analysis procedure described for the ESPaDOnS spectra is used here, though in the $K$-band the temperature search range is expanded to $T \in [3000, 4200]$ K.

We find a larger dispersion in the point estimates for the stellar parameters derived from the IGRINS data than those derived in the optical spectrum.  The most conspicuous trend is in the derived photospheric temperature as a function of wavelength shown in Figure \ref{fig:SingleTeffvsOrder}.  The photospheric temperature peaks at values of $\sim4200$~K in the short wavelength end of $H-$band and saturates at $<3500$ K at the long wavelength end of $H-$band.  The $K-$band shows even lower derived photospheric temperatures of $\sim3300$ K, or 700 K cooler than estimated from optical.  No single photospheric temperature can describe all the spectral lines present in the high resolution optical and IR spectra.  
Sources with such discrepancies have been seen previously, for example in Figures 4 and 5 in \citet{bouvier92}.  These discrepancies are circumstantial evidence for the detection of spectral features attributable to a second photospheric component.

\begin{figure*}
 \centering
 \includegraphics[width=0.95\textwidth]{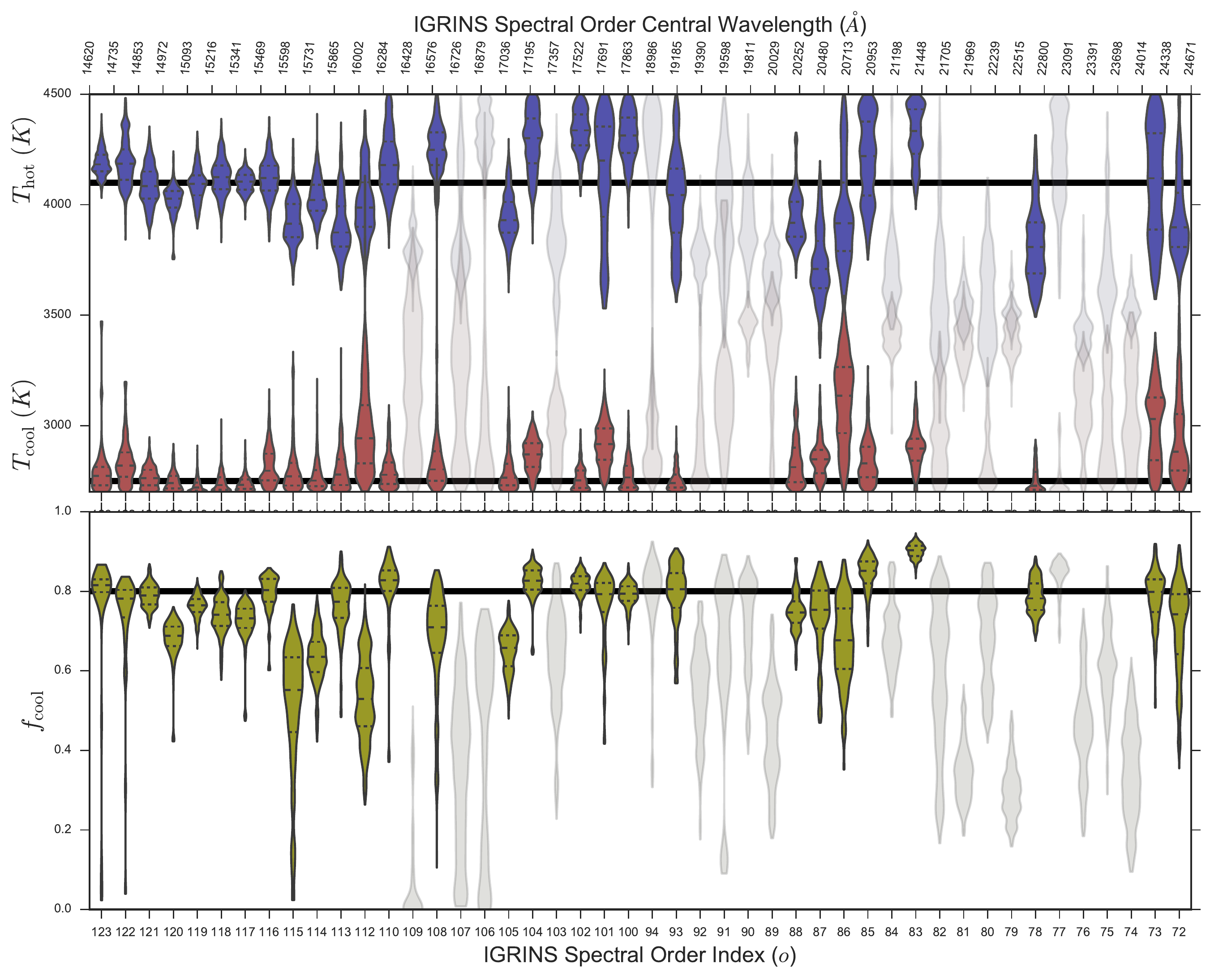} 
 \caption{Marginal probability distributions mirrored through the vertical axis \citep[``Violin plot'']{waskom14} for 48 IGRINS orders for $\teffa$ (blue shade in the top panel), $\teffb$ (red shade, top), and fill factor $f_{\mathrm{cool}}$ (yellow shade, bottom).  The stellar parameters are derived independently in each spectral order.  Spectral orders show differing levels of constraint on the cool photosphere and hot photosphere properties, including some orders ($o=104, 102, 100, 88, 83$) that yield especially tight cool spot filling factors.  The starspot temperature is consistent with values even lower than 2700 K, the lower limit of the temperature range used.  Many of the $K-$band orders that show lower estimates for the hot photosphere are unreliable due to spectral line outliers and uncorrected telluric residuals (light shaded distributions).}
 \label{fig:TwoTempResults}
\end{figure*}

\subsection{Heightened sensitivity to starspot spectral lines in the infrared}\label{sec:whyNearIR}

The heightened sensitivity to starspot spectral lines as a function of wavelength can be understood in the following way.  Starspots are cooler than their surrounding photosphere and will, therefore, have a longer wavelength of peak emission.  The average ratio of flux density between starspot and bulk photosphere will increase with wavelength until asymptotically approaching a fixed value on the Rayleigh-Jeans tail \citep{wolk96}.  The bottom panel of Figure \ref{fig:SingleTeffvsOrder} shows the flux density ratio for patches of 2800 K and 4100 K photospheres with equal areas (50\% filling factor).  The black-body ratios predict smooth flux ratio increases with wavelength, while the ratio of \PHOENIX\ models shows wavelength regions with heightened sensitivity to starspot fluxes, for example in $J-$band.  The visible portion of the spectrum will be dominated by the warm patches.  The infrared spectrum will have a lower overall spot contrast and will have more net starspot flux than optical emission.

Some care should be taken when directly comparing results between the ESPaDOnS and IGRINS spectra since they were not taken at the same time.  The particular ESPaDOnS spectrum used in Section \ref{sec:ESP_starfish} was obtained when LkCa 4 had an estimated $V-$band brightness of 12.94 mag, compared with 12.84 mag at the time of the IGRINS spectrum.  The fainter magnitude during the ESPaDOnS spectrum acquisition implies a greater coverage fraction of cool material than during the IGRINS spectrum acquisition.  
The photospheric temperature derived in the optical bands and short-wavelength end of $H$-band yield similar values of $\sim 4100$ K, suggesting the \emph{bulk}\footnote{The bulk appearance and disappearance of spectral lines is mostly controlled by the temperature of the emitting region of the local photosphere.  In other words, most of the variance in a spectrum is attributable to temperature, assuming near-solar metallicity. 
Starspots imbue dearths of flux in the line profiles of optical spectra, but these are secondary to the mere presence of temperature-sensitive lines.} spectral features are broadly consistent with emission from a single temperature component. A single ESPaDOnS order surrounding the TiO bands shows an exceptionally low estimated $\teff$.  The long wavelength portion of $H-$band and all of $K-$band are more sensitive to starspot spectral signatures than the shorter wavelength portions.

%%%%%%%%%%%%%%%%%%%%%%%%%%%%%%%%%%%%%%%%
% TABLE - History of LkCa4
%%%%%%%%%%%%%%%%%%%%%%%%%%%%%%%%%%%%%%%%
\begin{deluxetable}{ccc}

\tabcolsep=0.11cm
%\rotate
\tabletypesize{\footnotesize}
\tablecaption{Adopted values of \name from IGRINS spectra \label{tbl_adopted_props}}
\tablewidth{0pt}
\tablehead{
\colhead{$\teffa$} &
\colhead{$\teffb$} &
\colhead{$f_{\Omega}$}\\
\colhead{K} &
\colhead{K} &
\colhead{\%}
}
\startdata
   $4100\pm100$ &     $2750^{+250}_{-50}$ &    $80\pm5$ \\
\enddata
\tablecomments{These are the values at the epoch of the IGRINS spectrum acquisition.}

%\end{deluxetable*}
\end{deluxetable}

\subsection{Two-temperature fitting to IGRINS spectra}\label{sec:two_tempIGRINS}

Full-spectrum fitting was performed for 48 of the 54 available IGRINS spectral orders, omitting only the orders with the most pathological telluric spectral artifacts. We applied the mixture model as described in Section \ref{sec:methods} and Appendix \ref{methods-details}.  Standard MCMC convergence tests were evaluated.  The stellar parameter ranges were $\logg \in [3.5, 4.0]$ and $\Z \in [-0.5, 0.5]$; the \PHOENIX\ model spectrum temperature range was $\teffa, \teffb \in [2700, 4500]$.  $H-$band fits had normal distribution priors in place: solar metallicity to $\pm0.05$ dex, $logg=3.8\pm0.1$, and $\vsini=29\pm5$ km/s.

The fit quality is first assessed by examining the consistency of the point estimates of $v_z$ and $\vsini$ across the spectral orders.  The distribution of $v_z$ and $\vsini$ exposed extremely poor fits in two orders ($o=91$ and $94$), with all other orders demonstrating $v_z = 12.4 \pm 2.6$ km/s and $\vsini = 28.8 \pm 2.0$ km/s.

Overplotting forward-modeled spectra with the observed IGRINS spectra yielded insights on why some spectral orders perform better than others in assessing stellar properties.  Orders with extremely poor telluric correction residuals, large spectral line outliers, and uncorrected H line residuals from A0V standard division, are all excluded from our final stellar parameter compilation.  Several orders in $K-$band were rejected due to poor spectral fits influenced by deep metal lines.  Metal lines can be biased by Zeeman broadening, which could explain poorer fit quality with wavelength \citep{johnskrull07,deen13}.  Some orders were also discarded because their fits were mediocre or uninformative.  The remaining orders are relatively devoid of spectral line outliers and include the most information rich portion of the spectrum.

The IGRINS spectrum demonstrates some features that are present only in the hot photosphere model, and some features that are present only in the cool photosphere model.  Figure \ref{fig:specPostageStamp} shows a selection of six such spectral features for a range of plausible fill factors.  In some cases (\emph{e.g.} the top two panels), featureless cool photosphere spectra veil isolated spectral lines predicted in the hot photosphere models.  In other cases (\emph{e.g.} the middle row and lower left panel), the hot photosphere model veils sequences of shallow spectral features predicted in the cool photosphere model.  The cool spectral models predict shallow features because line blanketing from multiple indistinct molecular bands overlaps from rotational broadening; any feature of interest will be biased to non-zero veiling.  

The combination of line blanketing and high $\vsini$ has probably hampered efforts to identify isolated spectral features suitable for line-depth-ratio analysis in the spectra of rotationally broadened young stars.  One such feature, the OH 1.563 $\mu$m line analyzed by \citet{oneal01}, shows a clear pattern of 3 lines in our data, with the central line exceeding the depth of the adjacent two lines.  The \PHOENIX\ models predict a non-negligible hot photosphere contribution to the middle line for our range of hot and cool temperatures.  In comparison, the forward-modeling technique described in this work thrives in the presence of long sequences of indistinct yet predictably correlated spectral features.  The level of veiling of hot photospheric lines is set by the cool photosphere filling factor and temperature, while the level of veiling of the cool lines is set by the filling factor and temperature of the hot photosphere lines. Similar forward-modelling strategies have successfully identified patterns of weak metal lines in the line-blanketed spectra of low metallicity stars \citep{kirby11,kirby15}.  Figures \ref{fig:Hband3x7} and \ref{fig:Kband3x7} in the Appendix show 42 of the $H-$ and $K-$band spectra on a log scale with a single random composite model spectrum overplotted.  

Figure \ref{fig:TwoTempResults} shows the distribution of $\teffa$ (blue shading), $\teffb$ (red shading), and $f_{\mathrm{cool}}$ (yellow shading) for all of the spectral orders, with the rejected orders grayed out, and the reliable order subset shown in bold.  The point estimates for each spectral order are listed in Appendix Table \ref{tbl_order_results}.

Best fit values for the stellar parameters are listed in Table \ref{tbl_adopted_props}.  Remarkably, the filling factor of cool photosphere exceeds 50\%, with a best fit value of $f_{\mathrm{cool}}=80\pm 5 \% $.

\section{The Two Temperature Fit to the SED and Stellar Rotation}\label{sec:GJHsection4}

\begin{figure}
 \centering
 \includegraphics[trim=3.1cm 13cm 1.0cm 3.6cm, clip=true, width=0.49\textwidth]{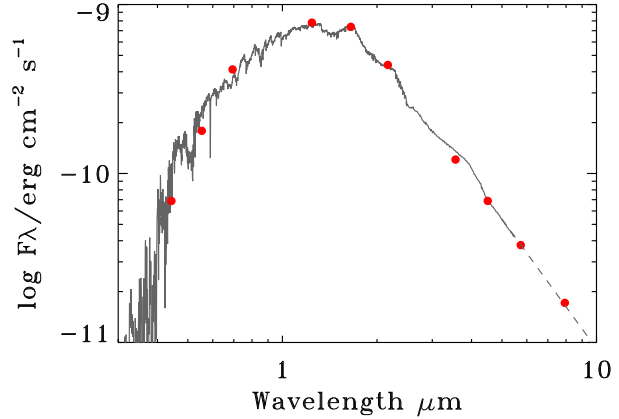}
\caption{The SED of LkCa 4 (red) compared with the two-temperature spectrum obtained from the best fit to the IGRINS spectra (black), scaled to $A_V=0.3$ and $\log L/L_\odot=0.20$ at the epoch of the 2MASS observation (see \S 5n).}
\label{fig:sed}
\end{figure}

\begin{figure*}
 \centering
 \includegraphics[trim=2.1cm 3.0cm 1cm 7.7cm, clip=true, width=0.70\textwidth]{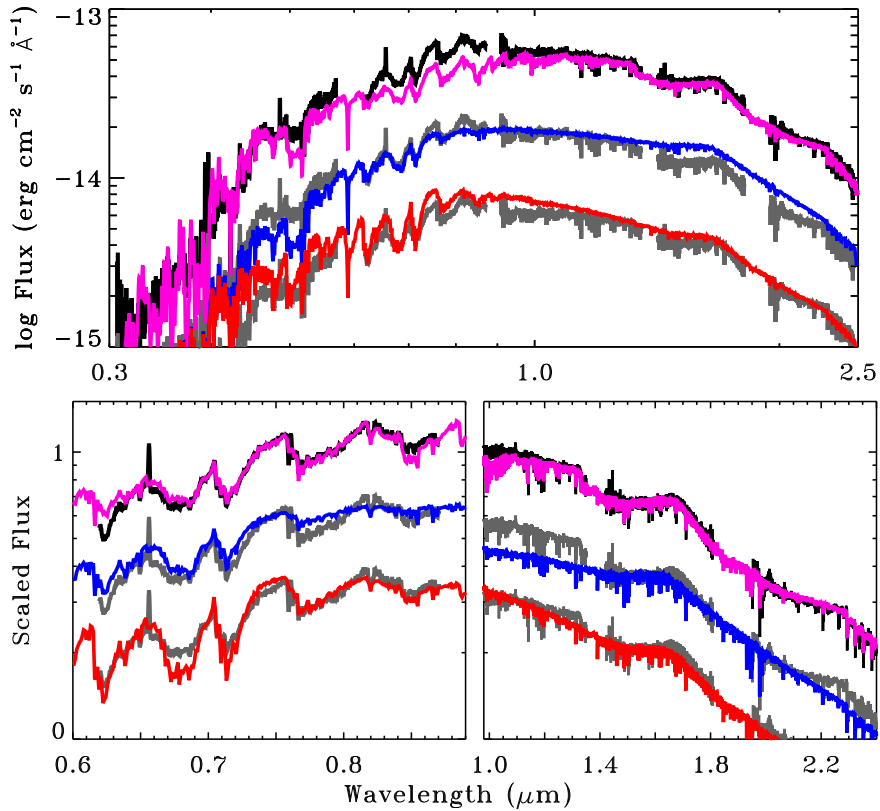}
 \caption{Top:  The low-resolution optical/near-IR spectrum of LkCa 4 obtained from Palomar/DBSP and APO/Triplespec on 30 December 2008 (black), compared to a synthetic spectrum of a two temperature photosphere (4100 K with 20\% fill factor and 2750 K with 80\% fill factor, reddened by $A_V=0.3$ mag; purple lines), a 3900 K spectrum reddened by 1.3 mag (blue lines), and a 3500 K spectrum with no reddening.  All synthetic spectra are fit to the $J$-band spectrum of LkCa 4. The observations in gray and the single temperature photospheres are scaled by factors of 3 and 9 for visual purposes.
 Bottom:  Same as the top panel, but the synthetic spectra are 
scaled separately to the optical spectrum at 0.75 $\mu$m and to the near-IR spectrum at 1.5 $\mu$m.  Warm photospheres accurately reproduce molecular bands at $0.7$ $\mu$m but fail to fit the spectral features at longer wavelengths.  Cooler photospheres predict molecular bands at $<0.7$ $\mu$m that are much deeper than observed.  The two temperature photosphere accurately fits spectral features in both optical and near-IR wavelengths.}
 \label{fig:lores}
\end{figure*}

\begin{figure}
 \centering
\includegraphics[trim=2.6cm 13.0cm 2.3cm 2.5cm, clip=true, width=0.48\textwidth]{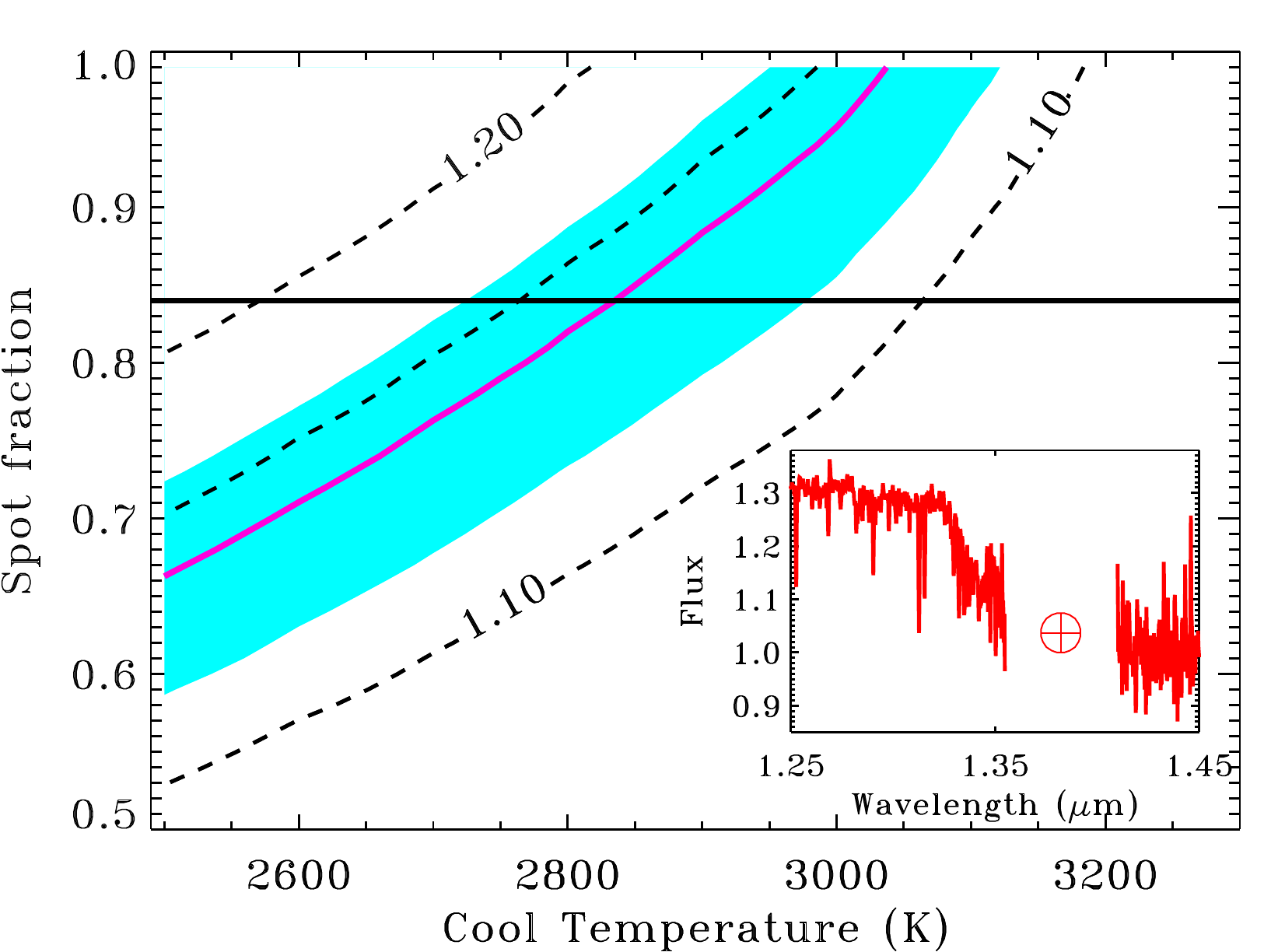}
\caption{The constraint on the size and temperature of the cool spot from the H$_2$O absorption band depth between $J$ and $H$-bands.  The contours show the flux ratio at 1.30 and 1.34 $\mu$m from the Phoenix models.  Comparisons to the APO/TripleSpec spectrum, shown in the inset and normalized at 1.4 $\mu$m, yield best-fit solutions (solid purple line) and the range of acceptable fits (shaded cyan region, with contours of flux ratio in dashed black lines) for an epoch when  the cool spot fraction was $\sim 84$\%.  The range of acceptable fits is based on an estimate uncertainty of $\sim 2$\% in observed flux ratio, which is dominated by uncertainty in telluric correction of strong H$_2$O absorption. }
 \label{fig:h2ojump}
\end{figure}

In the previous section, we established that the high resolution optical and near-IR spectra of LkCa 4 may be explained by a two-temperature photosphere.  In this section, we test the two component fit by comparing the combined spectrum to observed spectral features and rotational modulation.  We adopt the best-fit two temperature model to the IGRINS spectrum, with components of 2750 K covering 80\% of the visible stellar surface and 4100 K covering 20\% of the stellar surface, obtained when LkCa 4 had an estimated brightness of $V=12.84$ mag.  The cool and hot temperatures and the filling factor are blindly adopted without any adjustments to attempt to improve fits to the SED or broadband spectra.

Figures~\ref{fig:sed}--\ref{fig:lores} compare synthetic spectra from the two-component photosphere to the SED and flux-calibrated spectra of LkCa 4.  The only free parameters in this comparison, the luminosity and extinction, are both scaled to match the spectrum (see \S 5.1-5.2).  The optical-IR SED is obtained by estimating photometry from \citet{grankin08} during the 2MASS epoch, with $V\sim12.61$ mag (see Table~1).  The SED also includes mid-IR photometry from Spitzer/IRAC \citep{hartmann05}, without adjusting for epoch.  In the spectroscopic comparison, some minor discrepancies between the optical and near-IR spectra may be introduced because the spot coverage changed in the $\sim 5$ hrs between observations ($\Delta V=0.08$ mag).  The synthetic spectrum is obtained from the Phoenix models, as in \S 3, and is extended beyond the longest wavelength (5 $\mu$m) to calculate the bolometric luminosity.

The synthetic models match the full spectrum reasonably well.  When scaled separately to the red-optical and near-IR spectra, the synthetic models match even better (bottom panels of Figure \ref{fig:lores}), while single temperature photospheres fail to reproduce large spectral features.   The $3900$ K component fits the TiO bands at $<7400$ \AA\ but fails to reproduce molecular bands at longer wavelengths.  The $3500$ K component suffers from the opposite problem, yielding molecular bands at short wavelengths that are too deep.  The two temperature photosphere with the parameters from the best-fit calculated in \S 3 reproduces the TiO bands, the hump in the $H$-band, and the jump in flux at the long-wavelength end of the $J$-band.  This latter feature is caused by opacity in H$_2$O bands and is only seen when the fraction of emission from the cool component is high (Figure \ref{fig:h2ojump}).

These spectral features should be rotationally modulated as the filling factors of the visible hot and cool components change.    In \S 4.1, we first use our fit to the IGRINS spectrum to convert the $V$-band brightness to the filling factor of the two components.  We subsequently calculate the color changes expected based on this spot coverage and compare these results to historical data (\S 4.2), and confirm our basic approach by demonstrating that TiO band depths vary with rotation (\S 4.3).

\begin{figure*}
 \centering
 \includegraphics[trim=1.6cm 12.6cm 2.2cm 2.4cm, clip=true, width=0.60\textwidth]{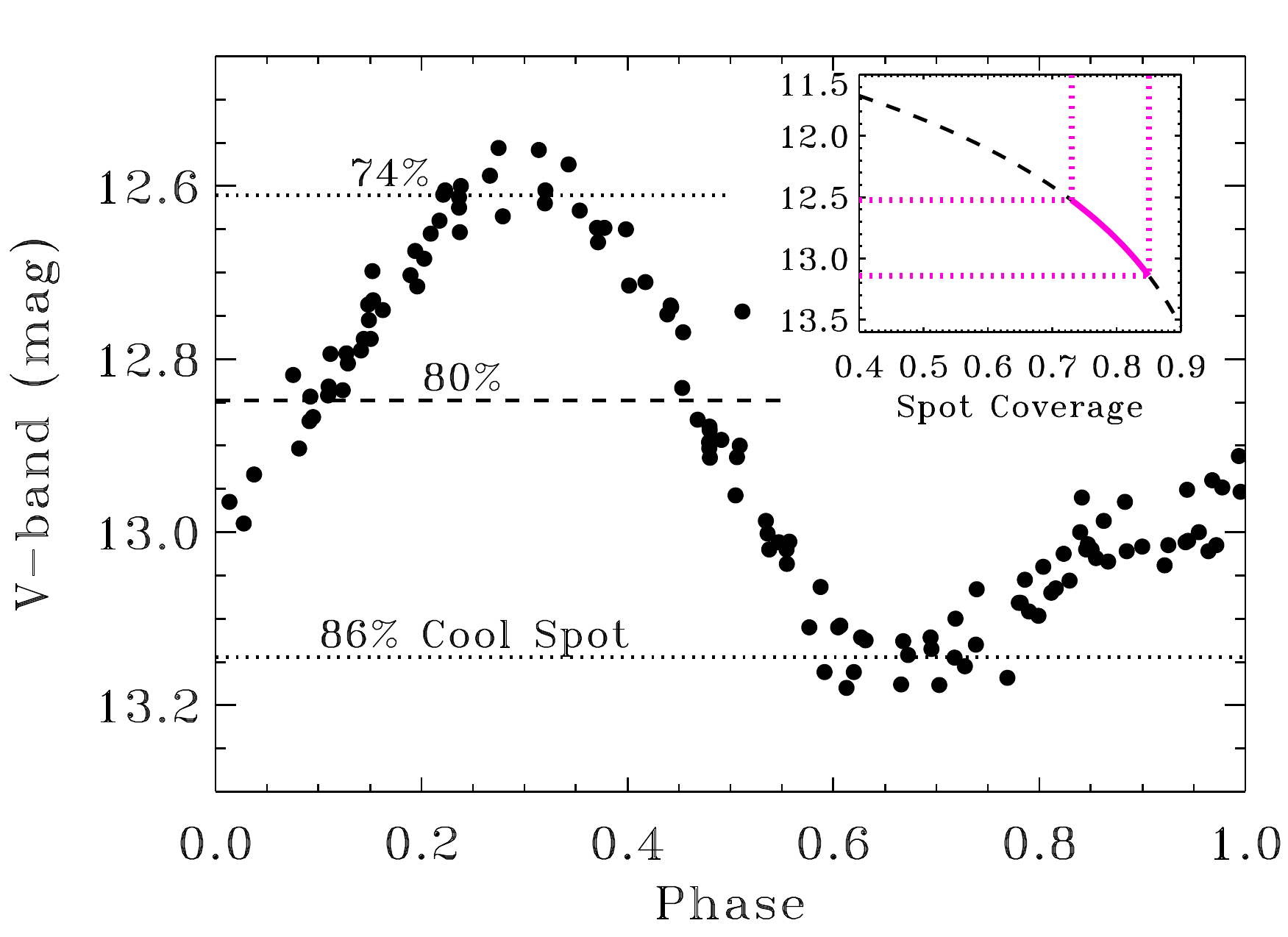}
\caption{The $V$-band magnitude in 2014--2015, converted into fill factor for the cool component.  The optical brightness depends mostly on the hot component.  If we fix a 75\% filling factor, as measured in the IGRINS spectrum, to $V=12.83$ at the time of the observation, then the $V$-band amplitude  corresponds to filling factors of 67--83\%.  The factor of $\sim 2$ change in visible surface area of the hot component, from 33\% to 17\%, is required to produce the $\Delta V=0.6$. }
\label{fig:vband_spot}
\end{figure*}

\begin{figure*}
 \centering
 \begin{tabular}{ll}
 \includegraphics[trim=2.6cm 13.0cm 2.8cm 3.2cm, clip=true, width=0.45\textwidth]{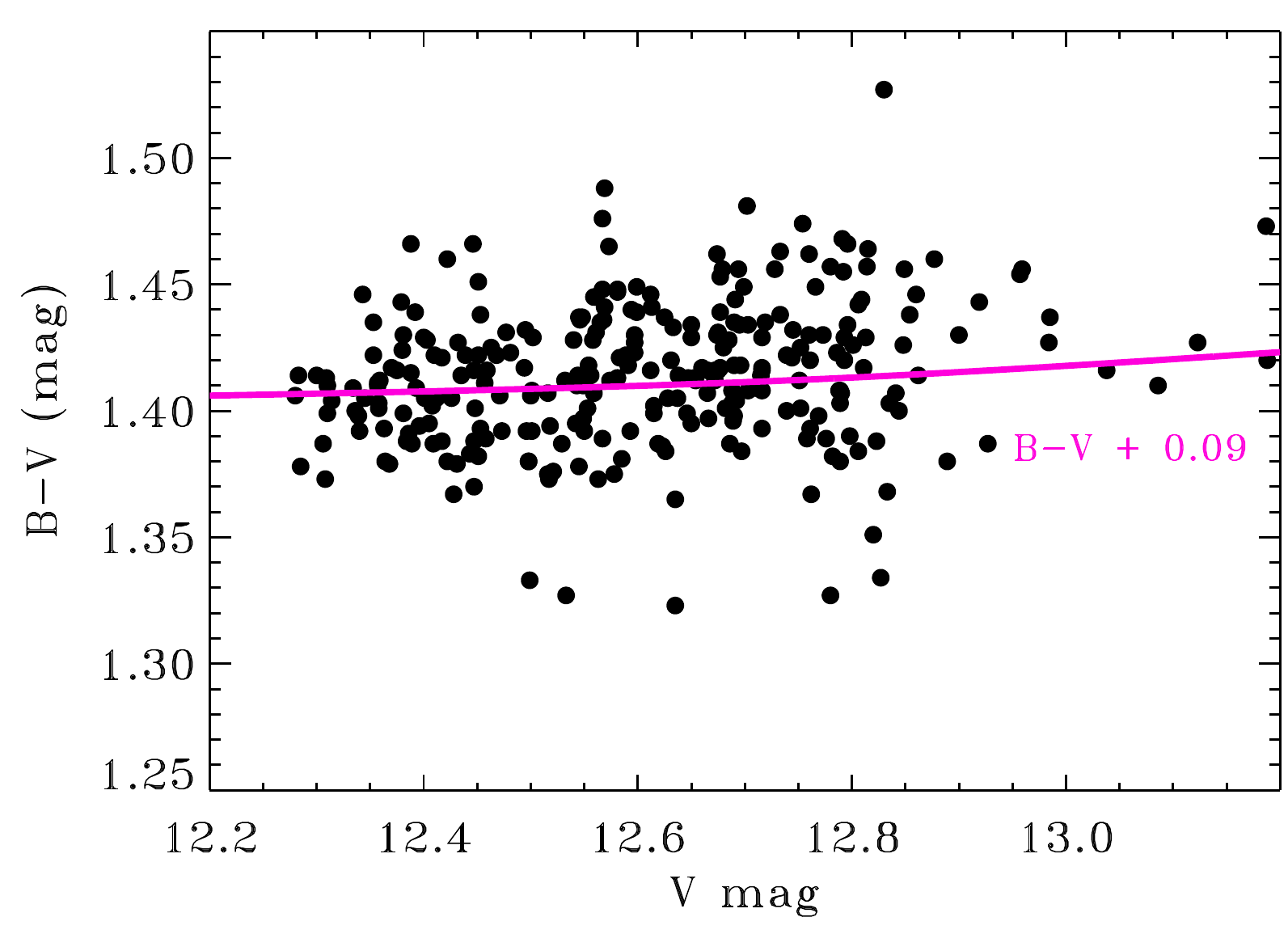} 
    &
   \includegraphics[trim=2.6cm 13.0cm 2.8cm 2.2cm, clip=true, width=0.45\textwidth]{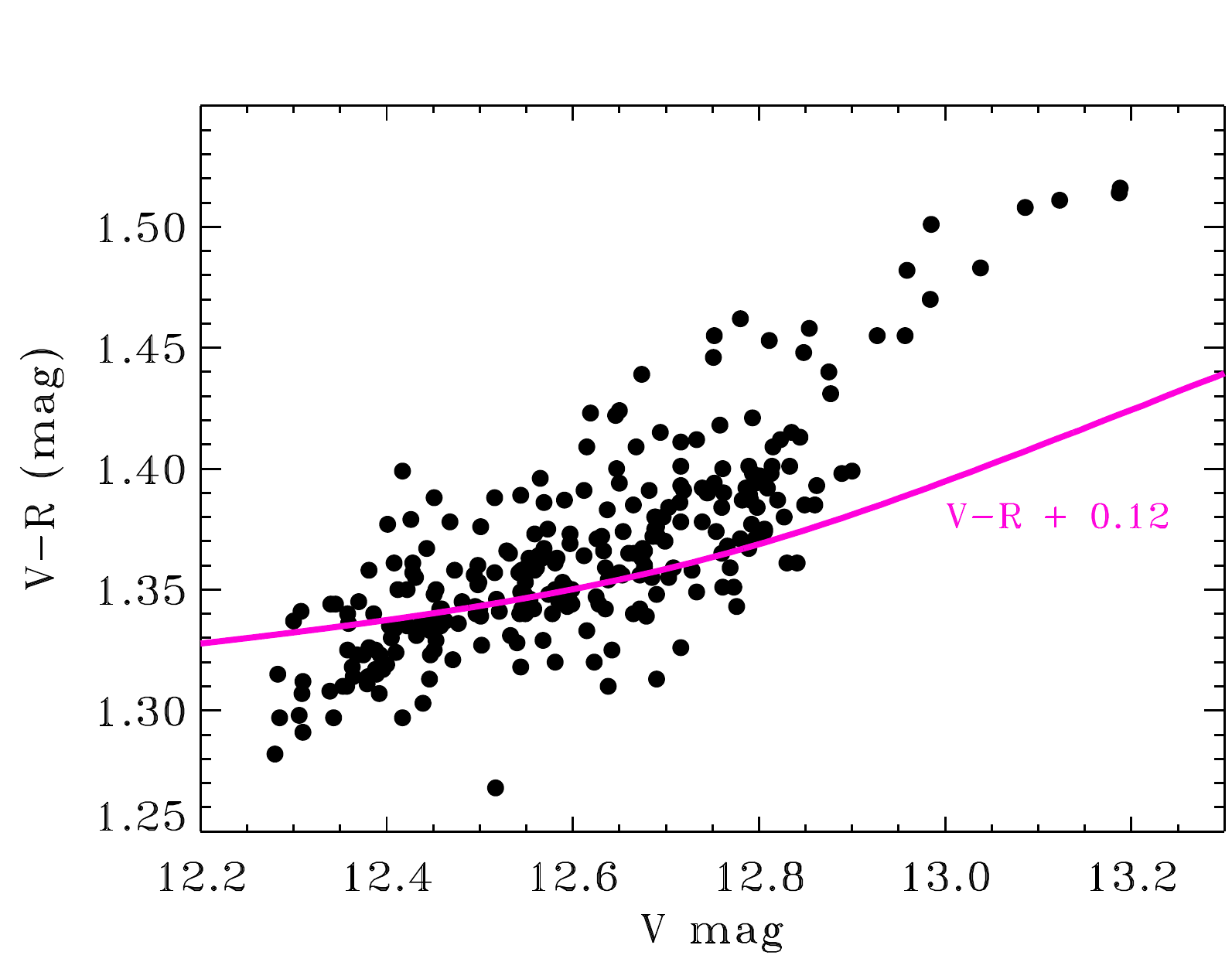}
    \end{tabular}
\caption{The observed optical colors of LkCa 4, mostly from \citet{grankin08}, compared with predictions from the rotation of a two-temperature photosphere model.  The model is constructed instead by converting $V$-band brightness to a cool spot filling factor and subsequently calculating colors from main sequence colors and bolometric corrections of \citet{pecaut13} (purple lines).}
\label{fig:colors}
\end{figure*}

\subsection{The $V$-band lightcurve and spot coverage}\label{sec:rotSpot1}

The  $V$-band brightness reflects the instantaneous filling factor of the cool and hot components.  In the two-temperature photosphere, the $V$-band emission is dominated by the hotter component and is, therefore, a good proxy for the visible surface area of the hot component.  Figure~\ref{fig:vband_spot} shows the winter 2015--2016 ASAS-SN lightcurve, converted into a cool spot filling factor.  During this period, the brightest and faintest phase corresponds to cool component fill factors of $74\%$ and $86\%$, respectively.  Since 1992 the $V$ band photometry has been as bright as $12.3$ mag, which corresponds to a cool component fill factor of $65\%$, and as faint as $13.2$ mag, or a cool component fill factor of $87\%$.  This drastic change in visible area of the hot spot (35\% to 13\%) is needed to explain the full $\sim 1$ mag range in brightness, assuming no temperature change in either component.

\subsection{Rotational Modulation of Colors}\label{sec:rotSpot}

The relative contributions from the hot and cool components change as LkCa 4 rotates, leading to color changes.  We model this rotational modulation using the main sequence colors \citep[compiled by][]{pecaut13}, which should have less contribution from spots than pre-main sequence colors.  The Cousins $R$ photometry from \citet{pecaut13} is converted to Johnson $R$ for comparison with the \citet{grankin08} photometry, following the color transformation prescribed by \citet{landolt83} and applied by Grankin et al.  Photometry from synthetic spectra produced by Phoenix and BT-Settl models cannot be used here because the predicted $B-V$ colors are much bluer than observed.  Offsets between model and observed colors are at least in part attributed to extinction and are discussed in \S 5.1.  

Limb darkening effects also become important if starspots are distributed in a finite number of large patches, not many small patches.  The variations in the mean projection angle cause confounding color trends in the photometric modulations.

The optical emission is dominated by the hotter component while both components contribute equally to the infrared emission (see inset in the top panel of Figure \ref{fig:lores}).  Figure~\ref{fig:colors} demonstrates that the $B-V$ color from \citet{grankin08} does not depend on $V$, which is consistent with expectations for a single hot component with no contributions from the cooler component.  The standard deviation of 0.03 mag in $B-V$ color (which includes a $\sim 0.01$ mag uncertainty in both $V$ and $B$) indicates that the spot temperature is stable to $\lesssim 50$ K over decades.  The stability of this temperature implies that the hotter component may be the ambient temperature of the photosphere.

The correlation between $V$ and $V-R$ indicates that the star becomes redder when the cool spot has a higher filling factor.  Our simple model predicts that the correlation should be much weaker than is measured.  Most likely, our two temperature model for the photosphere is overly simplistic, since the $V-R$ color could be reproduced with some contribution from intermediate temperatures.

Rotational modulation is expected to lead to much smaller brightness changes at near-IR wavelengths, with amplitudes of $0.15-0.2$ mag, than the $0.6-0.8$ mag amplitudes seen at optical wavelengths.  The smaller amplitude of near-IR brightness is also seen in a few spotted stars in the optical/IR monitoring of NGC 2264 \citep{cody14}, although those stars have a much smaller $V$ band amplitude than LkCa 4.

\subsection{Rotational Modulation of TiO Band Depths}\label{sec:RotTiO}

\begin{figure*}
 \centering
 \includegraphics[trim=3.2cm 3.3cm 2.6cm 8.3cm, clip=true, width=0.65\textwidth]{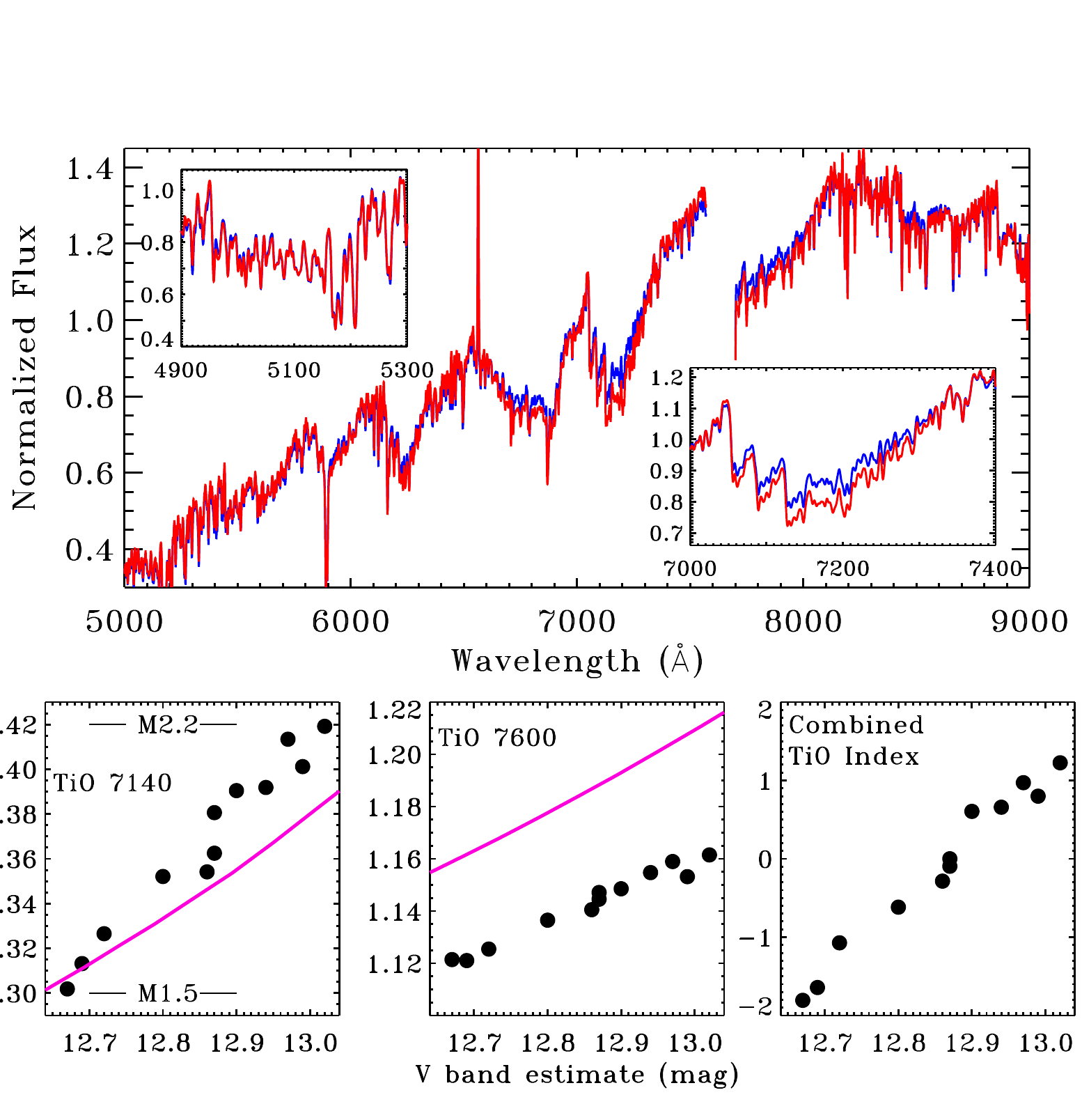}
 \caption{Variability in TiO bands measured with ESPaDOnS spectra.  The top panel compares spectra between when LkCa 4 was bright (blue) and faint (red).  The $V-$band emission is estimated from fits to the ASAS-SN lightcurve obtained during the same period.  The bottom left and middle panels show correlations between $V-$band magnitude and individual TiO band indices, while the bottom right panel shows a similar correlation with the average of the TiO-6200, CaH-6800, TiO-7140, TiO-7600, and TiO-8500 indices.}
 \label{fig:tiovar}
\end{figure*}

The depths of TiO and other molecular bands are common diagnostics of spectral type and therefore temperature in optical spectra \citep[e.g.][]{kirkpatrick91}.  In a 2-temperature photosphere cool enough for molecules to form, the depth of observed TiO bands will depend on the fractional coverage of the components.  The TiO band depths vary with spot coverage using 12 epochs of CFHT/ESPaDOnS spectra obtained over 14 days ($\sim 4$ rotation periods) by \citet{donati14}.

The spectra obtained near the estimated maximum and minimum optical brightness are shown in the top panel of Figure~\ref{fig:tiovar}.  Small but significant changes are seen in the TiO bands.  On the other hand, the blue spectrum does not change.  The TiO band depths in these spectra are measured from spectral indices as described in \citet{herczeg14}.  The TiO indices are also combined into a single TiO index by calculating the number of standard deviations each point is from the median value of the index and then averaging the standard deviation.  
% Could use a footnote to be more exact on the TiO indices.

The TiO bands trace the cool component.  When the hotter component has a larger filling fraction, the TiO bands are veiled by the hotter component and should, therefore, be shallower. As expected, 
the TiO indices all correlate strongly with the optical brightness.  The TiO-7140 band depths are accurately reproduced by varying the two-temperature fit with rotation.  Other bands have small offsets (see middle panel of Figure~\ref{fig:tiovar}), likely caused by either contributions from intermediate temperatures or by small errors in the synthetic spectra.  In the spectral type scheme derived by \citet{herczeg14}, the change in TiO-7140 index corresponds to a range from M1.5 (3640 K) to M2.2 (3530 K).

\section{Placing LkCa 4 on the HR diagram}

\subsection{Extinction estimates to LkCa 4}

An accurate extinction estimate from the full SED is challenging because of a strong dependence on the distribution of temperatures in the photosphere, which in our work is simplified to a two temperature fit.  However, isolating specific colors constrains the possible range of extinctions.  Based on the following analysis, an $A_V=0.3$ mag is adopted throughout this paper, although some uncertainty is introduced by confusion in the choice of template colors.

The $B-V$ color is stable and not affected by cooler components, and may therefore be compared with the expected colors for the hotter (4100 K) temperature.  
The average $B-V=1.41$ mag color from \citet{grankin08} has a color excess $E(B-V)=0.09$ mag relative to the empirical color expected for a main sequence star and $E(B-V)=0.20$ mag for colors of pre-main sequence stars, based on the temperature-color tables of \citep{pecaut13}.  The main sequence colors are probably most appropriate in this analysis because they are freer of spots.  
The $B-V$ color excess leads to an extinction $A_V=0.28\pm0.15$ mag relative to main sequence colors, or $0.62\pm0.15$ mag relative to pre-main sequence colors.   The listed uncertainties are calculated from extinction estimates relative to colors of 4000 and 4200 K photospheres.
The $V-R_J$ color excess leads to $A_V=0.48$ mag, which should be less affected by gravity (based on $V-I_C$ colors in Pecaut et al.) but is likely to be overestimated because the cool component contributes to the $R_J$-band emission, as seen in Fig.~\ref{fig:colors}.
In the near-IR, the estimated $V=12.61$ mag at the time of the 2MASS observations leads to a spot-weighted template (main sequence) color 
$J-K_s=0.87$ mag, with a range of $\sim 0.05$ mag for acceptable solutions with other temperatures and covering fractions. The observed $J-K=0.93$ mag leads to $A_V=0.35$ mag, 
with a range of $0.29$ mag to account for the range of other acceptable photospheric solutions.  
The template near-IR color is more robust than optical colors to differences between the main sequence and pre-main sequence temperature scales because the pre-main sequence $J-K$ color is redder at 4100 K and bluer at 2750 K.  

The near-IR extinction to LkCa 4, and perhaps to other TTSs, will be overestimated if emission is generated from a multi-temperature spectrum but template colors are based on only the hotter component.  This mismatch likely explains why near-IR extinctions are often higher than optical extinctions for young stars \citep{herczeg14}, IR colors that are often redder than expected from models \citep{tottle15}, and general challenges in near-IR templates of T Tauri stars \citep[e.g.][]{espaillat10,fischer11}.  For LkCa 4, the excess $E(J-K)$ color leads to an extinction of $0.84$ mag for a 4100 K photosphere and \citet{pecaut13} pre-main sequence colors.  This extinction is consistent with the color and SED analysis of \citet{furlan06}, where a K7 spectral type was adopted, and is higher than extinctions of $A_V=0.69$ mag measured from optical photometry \citep{kenyon95} and $0.35$ mag measured by comparing optical spectra to young spectral templates \citep{herczeg14}.

Our results tentatively support the lower extinction obtained at optical wavelengths, although the template colors used to calculate extinction are uncertain and will need revision.  Differences between main sequence and pre-main sequence temperature scales and colors in the \citet{pecaut13} tables are introduced by a combination of spots, in addition to any color differences caused by gravity differences.
Uncertainty in colors may be solved with self-consistent comparisons to young stars with similar spectral types if spot coverage is also similar.  Discrepancies in colors, spectral type/effective temperatures, and extinctions for LkCa 4, and perhaps more broadly for TTSs, are dominated by ignoring spots and not by spectral type-effective temperature scales.  Fundamental problems in molecular opacities and photospheric models are likely important secondary effects.

\subsection{Spot-corrected HR diagram placement}

\begin{figure*}
 \centering
 \includegraphics[trim=2.6cm 13.0cm 2.8cm 3.5cm, clip=true, width=0.65\textwidth]{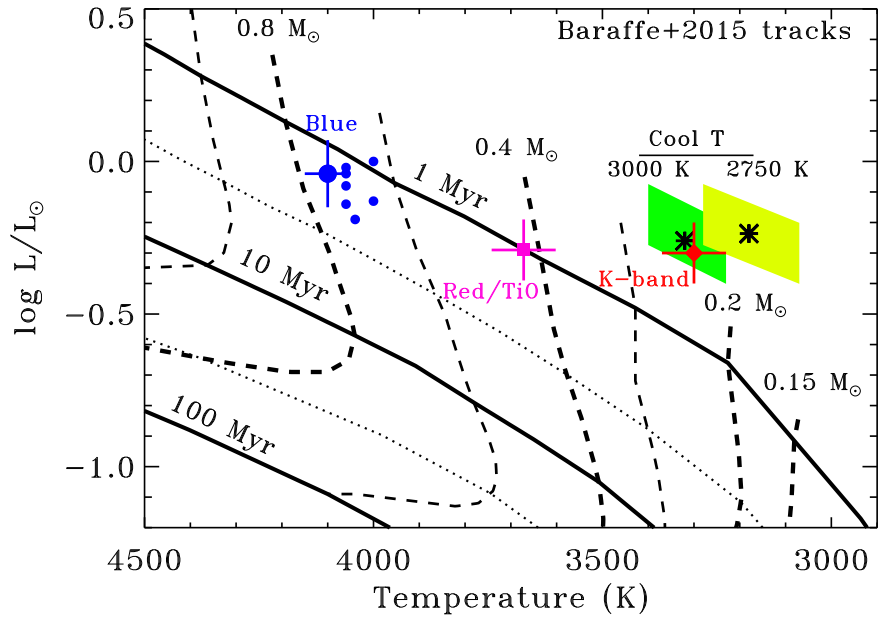}
 \caption{Locations for LkCa 4 on an HR diagram, compared with models of pre-main sequence evolution calculated by \citet{baraffe15} with isochrones (black lines) and evolution models of a single mass (dashed blue lines) as marked.  The measured effective temperature and luminosity from this paper, based on the two-component fit and a median $V$-band magnitude, is shown as the black asterisk.  The yellow shaded region corresponds to the range of apparent effective temperatures that would be measured as the hot component rotates into and out of our view for a cool temperature of 2750 K, while the green shaded region shows the same range for a cool temperature of 3000 K.  The blue circles correspond to the measurement at blue-optical wavelengths (the large circle and error bar is from \citet{donati14}), the purple square corresponds to the measurement from low-resolution optical spectra, biased to TiO band strengths, by \citet[][biased to TiO bands]{herczeg14}, while the red diamond corresponds to what we would measure from the K-band spectrum and 2MASS $J$-band magnitude.}
 \label{fig:hrdiag}
\end{figure*}

The spectrum of LkCa 4 is well reproduced by a cool component of $2750$ K covering 80\% of the surface and a $4100$ K component covering 20\% of the surface.   Scaling this composite spectrum of LkCa 4 to the observed 2MASS photometry ($\sim 74\%$ spot covering fraction) yields $R=2.3$ $R_\odot$ and $\log L/L_\odot=-0.20$, adopting $A_V=0.3$ mag and a distance of $131 \pm 2$ pc\footnote{This distance is consistent with the Gaia distance of $130 \pm4$ pc measured to the nearby Taurus members V410 Tau, V773 Tau, and BP Tau \citep{gaia2016dr}.} from the range of parallaxes of nearby Taurus members measured by \citet{torres12}.  
At the median 80\% spot covering fraction from 2015, the luminosity would be $\log L/L_\odot=-0.26$.

The effective temperature $T_{\rm eff}$ of these two components may be calculated as
\begin{equation}
T_{\rm eff}=[T_{\rm hot}^4(1-f_{\mathrm{cool}})+T_{\mathrm{cool}}^4 f_{\mathrm{cool}}]^{0.25}.
\end{equation}
For the median fill factor of 80\%, the observed effective temperature would be 3180 K, with a range of 3070--3280 for 86\% and 74\% spot fill factors.  If the cool spot is 3000 K, at the limit of the acceptable temperature range, then these values would increase by 150 K.  The apparent effective temperature in any single epoch depends on the relative fractions of the hot and cool spot.  When accounting for the visible contributions of both hot and cool components, this wide range of possible observed effective temperatures in any single epoch underscores the difficulty in placing young stars on HR diagrams.  

Figure~\ref{fig:hrdiag} and Table~4 show that this effective temperature is much cooler than previous measurements from optical spectra, which are clustered around 4000 K.  The luminosity does not change significantly because the broadband SED yields similar luminosities.

When the photosphere is assumed to be a single temperature, the effective temperature depends on the wavelength where the measurement is made.  For LkCa 4, the $K$-band temperature is closest to the effective temperature of the entire photosphere.  These discrepancies may also explain the large temperature, systematic offsets identified when comparing temperatures from $H$-band APOGEE spectra to those obtained from other (usually optical) methods \citep{cottaar14}.  Two component photopheres would also explain deriving different spectral types in the optical and infrared for the same source.
On the other hand, many GAIA-ESO spectra cover only a narrow wavelength around H$\alpha$ \citep[e.g.][]{jeffries14,frasca15}, which would yield an effective temperature of $\sim 4000$ K.   Spectral types and effective temperatures measured from TiO depths \citep[e.g.][]{herczeg14} are cooler than those estimated from blue spectra but still overestimate the effective temperature.  Temperature uncertainties for young K and M dwarfs are larger than any formal uncertainties in individual measurements.

In principle, the improved characterization of the photospheric emission and radius of LkCa 4 should lead to more accurate estimates of mass and age from pre-main sequence tracks.  However, most standard pre-main sequence stellar evolution models do not take into account the effects of starspots, and those that do \citep{somers15} limit cool spot areal coverage fractions to less than 50\%.  Comparison of LkCa 4 to \emph{unspotted} models is inherently flawed, but for the purposes of gauging the effect size of spots on derived parameters, our effective temperature and luminosity yield a mass of $0.15-0.3$ $M_\odot$ and an age of $\sim 0.5$ Myr in the \citet{baraffe15} evolutionary models\footnote{The \citet{baraffe15} evolutionary tracks were calculated using BT-Settl photospheric spectra \citep{allard14}, which may introduce minor uncertainties when compared to results from the PHOENIX models used here.}.  This age is much younger and this mass is much lower than previous estimates made assuming the warm component solely accounted for the effective temperature.

Even though LkCa 4 is only a single data point, the cool spot coverage fraction may be evidence that strong magnetic fields are inhibiting convection in LkCa 4.  As a result, the surface is cooler and releases less radiation, slowing the contraction rate.  LkCa 4 then appears more luminous and cooler than expected for a star of its genuine age from evolutionary tracks that do not consider magnetic fields.   This shift to lower temperature may also help to resolve some of the age discrepancies between intermediate mass and low mass young stars \citep[e.g.][]{herczeg15}.  The corrected placement of spotted stars on the HR diagram reflects the accurate luminosity, radius, and effective temperature, but evolutionary models are not yet equipped to interpret such heavily spotted stars \citep{somers15}.  In an HR diagram, spotted evolutionary model tracks at a given mass and age shift to higher luminosity and lower effective temperature.  Such a \emph{spot-corrected} HR diagram position would alter the age and mass estimates to $> 0.2-0.3$ $M_\odot$ and $\> 0.5$ Myr; in other words, back in the direction of previous na\"{i}ve estimates.  However, the full magnitude of the correction is an outstanding question.

The IGRINS and ESPaDOnS spectra indicate $\vsini\sim$ 28 km~s$^{-1}$ and a rotation period of 3.375 days. Combining these numbers gives $R\sin{i} \sim 1.87 R_{\odot}$. Our HR diagram analysis gives $R \sim 2.3R_{\odot}$, which suggests  $\sin{i} \sim 0.813$, or an inclination of about 35$^{\circ}$ from edge on. These values show broad consistency between rotational properties, spectral fitting values, and our interpretation of a tilted star revealing a circumpolar region with large areal coverage of starspots.

As seen in Figures \ref{fig:PhotTime} and \ref{fig:PhotPhase}, the areal coverage of spots appears to wax and wane secularly through the observing seasons of the last 31 years.  LkCa 4 is currently in a relatively faint phase, 0.2 mag fainter in $V$ than it was through much of the late 1980's and early 1990's.  It is conceivable that LkCa 4 happens to be going through a short-lived episode of relatively high coverage of cool photosphere analogous to--- albeit distinct from---solar maximum.  The intensity and duration of such episodes would further confound the interpretation of any instantaneous position on the HR diagram.

\section{Discussion}

In this paper, we have demonstrated that the IGRINS spectra of LkCa 4 are well fit with two temperatures, a hot photosphere of 4100 K and large cool patches of 2700--3000 K that cover 80\% of the stellar surface.  This solution reproduces the high-resolution $HK$-band spectra obtained with IGRINS, the broadband SED, and rotational modulation of spectral features.  Such a high covering fraction for a spot seems extraordinary, but may be common among young stars.  In this section, we describe that the need for a large cool spot is robust, despite some simplifying limitations of our spectral inference technique.  We also discuss how such a large spot could be interpreted, and suggest that even though LkCa 4 appears extreme, other young sources have similar characteristics.

\subsection{An absolute lower limit of starspot filling factor} 

This simple thought experiment demonstrates that the high starspot filling factor must be approximately correct.  A lower limit to the starspot filling factor can be set by the size of a black spot that can induce such large $\Delta V$.  Consider the following extremely conservative assumptions:  \begin{enumerate}
  \item The starspot emits \emph{zero flux}.  This statement is equivalent to the starspot being at \emph{absolute zero}.
  \item The stellar disk shows no starspots at maximum brightness.
  \item The starspot transits the center of the stellar disk.
  \item The star shows no feature brighter than the quiet photosphere at the visible surface
\end{enumerate}

For LkCa 4 in 2015 the $\Delta V\sim 0.5$ mag amplitude corresponds to an absolute minimum fill factor $f>0.37$.  At least 37\% of the stellar disk is covered in starspots at minimum light.  In the 2004 season, LkCa 4 exhibited $\Delta V\sim 0.8$, corresponding to a minimum fill factor of $f=0.52$.  If any of assumptions 1-3 is relaxed, the minimum fill factor \emph{increases}.  If assumption 4 is relaxed, the minimum fill factor decreases.  The large amplitude photometric variations in LkCa 4 are therefore evidence for either a hot spot in excess of the ambient photosphere or a large coverage fraction of cool starspots.

\begin{figure}
 \centering
\includegraphics[width=0.45\textwidth]{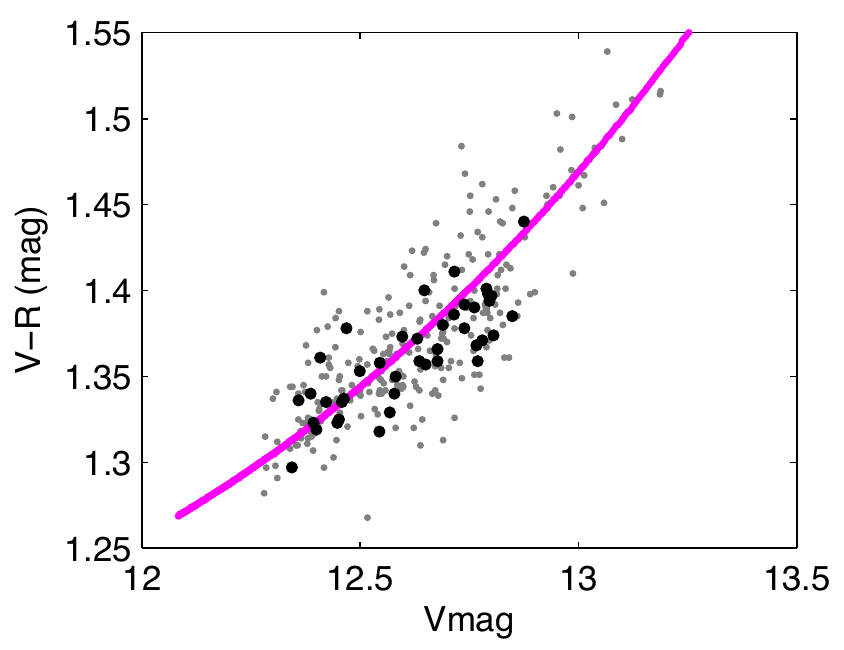} 
\caption{Grey points are all \citet{grankin08} photometric data, while black points are the subset of observations from the 1999 season.  The purple line represents a two component photometric model  described by \citet{grankin08}, comprising a photospheric temperature of 4040K (for K7), and a temperature of spotted area of 3030K (for an M5). In this model, the spot coverage of the observed hemisphere varies from $\sim36\%$ at maximum brightness, and $\sim82\%$ at minimum brightness.}
 \label{fig:grankin_vr}
\end{figure}

\subsection{Interpretation of the large filling factor of cool starspots}
We have shown that LkCa 4 exposes between 74\% and 86\% cool photosphere towards our viewing direction in winter 2015-2016, the most recent season for which there is data.  In previous seasons, like fall 2004 (\emph{c.f.} Figures \ref{fig:PhotPhase} and \ref{fig:vband_spot}), the filling factor of cool spots could have exceeded 90\%.  This exceptionally large fraction of surface area covered by spots defies the conventional wisdom that starspots are the minority constituent of the stellar surface.  The evidence for this derived fill factor has been shown in the SED, high-resolution spectral inference, direct spectral detection, and TiO variability modeling.  Perhaps the strongest evidence for the large fill factor of cool photosphere is the large amplitude of $V-$band modulation, which cannot be explained with small spots, even if a spot temperature of absolute zero is assigned to the cool regions.

One conceivable interpretation of the 74\%--86\% coverage fraction of cool photosphere is a stellar axis inclined towards our observing direction, with a large circum-polar spot covering much of the observable hemisphere, as described by \citet{donati14}.  The ambient hot photosphere would then be close to the equator, which would be limb-darkened due to projection effects.  Such limb darkening would cause a non-linear transformation between our observed fill factor $f_{\mathrm{cool}}$, and the surface areal coverage fraction of spots.  It is conceivable, and probably likely, that the value we report for $f_{\mathrm{cool}}$ is an overestimate of the surface area actually covered by spots.  In this circum-polar spot scenario, local fingers of cool spots could reach down to near the equator, entering into and out of the field of view and giving rise to the large amplitude photometric variation seen on LkCa 4.  This scenario is consistent with the inclination derived from relating $\vsini$, rotation period, and estimates for the stellar radius, as previously noted.  Such estimates are limited by estimates of the radius since uncertainties in effective temperature and luminosity cause a wide range of radius estimates.

The starspot geometry is also consistent with morphologies observed in interferometric maps of $\zeta$ Andromedae: large areal coverage of permanent polar spots, and smaller, transient networks of varying darkness which thread across the whole surface \citep{roettenbacher16}.

\subsection{Is \name an extreme source?}
The large amplitude of cyclical optical variability of LkCa 4 exceeds all other counterparts from the long-term monitoring of WTTSs by \citet{grankin08}.  From a statistical perspective, it might not be a surprise that the source with the largest photometric variability also possesses among the largest coverage fraction of cool spots.  When compared with spot covering fractions of 15--60\% for evolved giants and subgiants \citep{chugainov76,berdyugina05}, the 80\% spot coverage of LkCa 4 seems high.  However, photometric and spectroscopic evidence demonstrates that many other young low-mass stars also have large spot covering fractions.

Several WTTS in the Taurus-Auriga star-forming region have large amplitudes in periodic light variations, achieving $0.4-0.8$ mag in the $V-$band: LkCa 4, V410 Tau \citep{herbst89}, LkCa 7, V827 Tau, V830 Tau, and V819 Tau \citep{grankin08}.  In a sample that is not biased to variable stars, WTTSs observed in NGC 2264 show photometric variations similar to--but of lower average amplitude than-- LkCa 4 \citep{cody14,venuti15}.  Modeling of polychromatic photometric monitoring of the locus of WTTSs indicates a lower limit to the true coverage fraction $f_{\mathrm{eq}}\sim10-30\%$ and a difference $\teffa-\teffb\sim500$ K. 

Large amplitudes of the light variation indicate the existence of very extended spotted regions on the stellar surface. To estimate the total area and mean temperature of the spots in the visible stellar hemisphere Grankin used a simple non-parametric model for analysis of V410 Tau, V819 Tau, V827 Tau, and V830 Tau \citep{grankin98,grankin99}.  For the stars with the highest light curve amplitudes, this model shows that the spotted regions cover from 17 to 73\% of the visible stellar hemisphere and that the mean temperature of the cool regions is $500-1400$ K lower than the ambient photosphere.  An analysis of the calculated spot parameters of V410 Tau revealed a correlation between the amplitude of periodicity and the spot distribution
nonuniformity, which is defined as the difference of spot coverage in the visible hemispheres of the star at maximum and minimum light. As the amplitude increases from 0.39 to 0.63 mag, the degree of nonuniformity in the spot distribution increases from 21 to 37\%.  An alternative analysis of TiO bands indicate a spot coverage of $\sim 41$\% for V410 Tau \citep{petrov94}.  Besides the high amplitude of variability, these objects show the phenomenon of long-term stability of a brightness minimum in the interval from 5 to 19 years.

The Pleiades provides an older \citep[$\sim125$ Myr,][]{stauffer98} sample of young stars, many of which have starspots that cover large fractions of the stellar surface.  For example, anomalous colors of K dwarfs in the Pleiades can be explained by $\sim$50\% fill factors of spots \citep{stauffer03}, while LAMOST spectra of the TiO band indicates that $\sim 30$\% spot coverages may be common for 3500--5000 K stars \citep{fang2016}.  Recent photometric monitoring of the Pleiades indicates that the anomalous colors scale with rotation rate, strengthing the case for magnetic-field induced starspots as the cause of the anomaly \citep{covey16}.  However, the \citet{covey16} Pleiades rotation rates do not scale with the $\Delta V$ photometric amplitude at the precision of the PTF data, suggesting that while the overall starspot coverage increases with rotation rate, the longitudinally asymmetric component does not.  LkCa 4 represents an example of such an architecture, with more than  50\% of the visible stellar surface covered in spots at all observational phases. The longitudinally symmetric component evades photometric modulation detection but contributes to otherwise anomalous color offsets.

Spots are also likely important in estimating the properties of generic low-mass young stars. As an example, the well-studied CTTS TW Hya has a $\sim 4100$ K photosphere measured from high-resolution optical spectra \citep[e.g.][]{yang05}.  However, the near-IR spectrum is consistent with $\sim$ M2, when compared with main sequence dwarf stars \citep{vacca11}.  \citet{debes13} reconciled this discrepancy by invoking spots, although the spectral contribution from accretion and disk continuum emission was uncertain.  \citet{mcclure13} found that the TW Hya spectrum was well reproduced with near-IR spectra of stars with similar ($\sim$K7) optical spectral type, as long as the comparison spectra were young.   \citet{herczeg14} found an intermediate spectral type for TW Hya and other TTSs by focusing on TiO bands at redder wavelengths than previous spectra, and also by using young stars as photospheric templates.  Similar discrepancies have been found between optical and near-IR spectral types \citep[e.g.][]{bouvier92,bary14,cottaar14} and in comparisons of spectral types measurement based on TiO bands and those based on high-resolution blue spectra \citep[see discussion in][]{herczeg14}.

The mismatches between near-IR spectra of dwarf stars and young stars (along with some of the color discrepancies, see Section 5.1) is also likely explained by the prevalence and importance of cool spots in low mass stars.  For classical T Tauri stars, the veiling at $\sim 1$ $\mu$m is usually larger than can be explained by either most accretion models\footnote{\citet{ingleby13} invoked multiple accretion flows with a range of densities to explain this red excess.} or warm disk emission \citep[e.g.][]{basri90,cieza05,fischer11,mcclure13}.  Similarly, \citet{cottaar14} found that most cool stars in the Pleiades are veiled in the H-band, with no possibility for any disk or accretion emission.  Differences in spot temperatures and coverages likely contribute to and may explain entirely this excess veiling.

\subsection{The limitation of a two-temperature model}

We assume only two temperature components are present in the photosphere.  In reality, starspots are probably described by a range of spot temperatures analogous to umbra, penumbra, and plages on the solar surface.  A model possessing three unique temperatures and two unique fill factors would better approximate the minor constituents of the emergent spectrum than our two temperature model can, but additional components will always appear to fit data better while leading to possible overfitting.  The fit-quality of the existing two-temperature fits seems strong through much of $H-$band and many $K-$band orders, so the contribution of minor photospheric constituents is relatively low.  Interestingly, the $K-$band has several spectral orders that converge towards hybrid temperatures in the vicinity of $\sim3500$ K.  Most of these orders are flagged as erroneous due to strong metal line spectral outliers or uncorrected calibration artifacts.  A few orders just show mediocre fits where no permutation of hot and cool two-temperature models could describe all of the variance in the spectrum.  These $K-$band orders could be examined for a penumbra-like temperature component.  We caution that systematic errors in the models would ultimately limit the interpretability of models with more than two temperature components.  Space-based high-precision planetary transit spectrophotometry could offer an avenue to measure the morphologies and temperature distributions of starspots for the subset of transiting exoplanet host stars or eclipsing binaries.

Analyses of optical emission also suggest that a two-temperature model may be an oversimplification.  Figure \ref{fig:grankin_vr} shows that the $V-R$ colors from \citet{grankin08} are better reproduced if the cooler component is $\sim 3050$ K, thereby contributing more flux at optical wavelengths.  The ZDI imaging of \citet{donati14} also indicates a complex structure of spots that are hotter than what is interpreted here as the ambient photosphere, while also reproducing the optical photometry.  Despite the need for more complicated emission structures to simultaneously reproduce all observations, the broadband SED, spectral shape, and rotational modulation of spectral indices are reasonably well fit with the two components presented here.  The high and low-resolution near-IR spectra, including the H$_2$O band absorption between the $J$ and $H$ bands (see Figure~\ref{fig:h2ojump}), require a cool (2700--3000 K) component while the global photometry requires a large covering fraction for this cool component.

\subsection{Limitations of assumptions in the spectral inference methodology}

We assumed that the \PHOENIX\ synthetic spectra are a good representation of the observed IGRINS spectrum of LkCa 4, a young star that possesses a strong magnetic field \citep{donati14}.  Cool sunspots are associated with locally heightened magnetic fields on the sun, so synthetic spectral models employing negligible magnetic fields will fail to fit magnetic sensitive features.  Magnetic fields will have a few effects on the emergent spectrum- line broadening attributable to the Zeeman effect, and pressure broadening attributable to seeing deeper into the photosphere in the line wings than in the line center.  The former effect will preferentially impact lines with high magnetic sensitivity \citep[\emph{e.g.}][]{johnskrull99,deen13}.  The latter effect could masquerade as an apparent shift in surface gravity, as seen in mixture models of spotted stars with empirical templates composed of dwarfs and giants \citep{oneal96}.  Zeeman amplification can also be present \citep{basri92}.

Our tacit assumptions have been that these line-broadening effects are 1) relatively small compared to the gross appearance and disappearance of spectral lines at disparate temperatures, 2) many orders and lines are not affected at all, and 3) the Gaussian process noise model provides some resilience against unknown residual spectrum correlations induced by such under-fitting.  We see tentative evidence for physics outside of the pre-computed model grids in the poor fits to the observed spectra in $K-$band.  Strong metal lines demonstrate large departures in their best fit $\Z$, as the na\"{i}ve fitting strategy futilely attempts to minimize the residuals from Zeeman-broadened lines.  No pre-computed model grids including a range of surface magnetic fields are available at the wavelength range and spectral sampling needed to approach these questions within this spectral inference framework.

Uncertainties in the \PHOENIX\ synthetic spectra continuum opacity flow down to uncertainties in the absolute value of the fill factor.  Erroneous bulk opacities could be responsible for some of the scatter of fill factor distributions seen across different spectral orders, as in Figures \ref{fig:TwoTempResults} and \ref{fig:specPostageStamp}.  The large number of spectral orders over which the fill factors are averaged offers resilience against these anomalies.

\subsection{Evidence for intra-spectrum RV jitter}
Starspots cause significant jitter in radial velocity (RV) surveys for young planets \citep[e.g.][]{donati14, robertson14}.  The two-temperature model assumed that the hot and cool components shared the same $\vsini$ and $v_z$.  This assumption is accurate for either homogeneously distributed spots, azimuthally symmetric bands of spots, or polar spots.  LkCa 4 has clear optical line-profile variations \citep{nguyen12, donati14}.  We should be able to detect an \emph{anticorrelation} between the line center positions of spectral lines arising from the ambient photosphere and lines arising from cool starspots, since there is a zero-sum competition for solid-angle on the recessional and advancing stellar limbs.  We found tentative, albeit weak, evidence for such anticorrelation in $v_z$ across spectral orders dominated by cool photosphere and those dominated by hot photosphere.  For example, the cool-photosphere dominated spectral orders $100, 102, 104$ ($1.7$--$1.8$ $\mu$m) show the largest median $v_z\sim14.5-17.0$ km/s, while the other orders that possess a mix of primarily hot- and some cool- photosphere features demonstrate $v_z\sim9-13$ km/s.  An independent analysis of the RV of the IGRINS spectrum yields a barycentric RV of $14.6\pm0.2$ km~s$^{-1}$.  The direct emission of starspots could constitute a heretofore unaccounted source of correlations in conventional RV signal processing.

Radial velocity (RV) planet searches in the infrared--where the modulation of RV signals is lessened \citep[\emph{e.g.}][]{prato08,crockett12}-- could potentially reduce RV jitter by characterizing and constraining the effect of direct emission from starspots, in addition to the flux deficits imbued in optical line profiles.  Direct spectral emission from cool starspots is likely much weaker in sources with less starspot coverage than LkCa 4.  As demands on RV precision increase, the characterization and modeling of direct emission from starspots should in principle decrease RV jitter attributable to starspots in the near-IR spectra of any spotted star.

\section{Conclusions}

This paper establishes that the optical and near-IR spectra of the weak-lined T Tauri star LkCa 4 are produced by multiple temperature components.   When the temperature distribution is simplified to two temperatures, the photospheric model that best fits the spectrum includes a hot (4100 K) photospheric region that covers 20\% of the stellar surface and a cool (2700--3000 K) photospheric that covers 80\% of the stellar surface.  The previously-established $V$-band brightness variability of $0.8$ mag is caused by the rotational modulation of the covering fraction of the hot photospheric component.  This spectral modeling leads to an improved effective temperature measurement of LkCa 4. For LkCa 4 and other young low-mass stars, the standard approach of inferring masses and ages of young stars is inappropriate and will lead to large uncertainties, both because effective temperatures are systematically too hot and because most models of pre-main sequence evolution neglect spots.

New results from \citet{fang2016} show that large starspot coverage fractions of $50\%$ are common even among stars in the Pleiades, which at $\sim 125$ Myr are much older than the 1--10 Myr old stars in the Taurus Molecular Cloud.  LkCa 4 breaks the apparent ceiling at 50\% coverage fraction of cool starspots, though it is not clear how to classify stars for which the minority of the surface is purportedly the ambient photosphere.  A high fraction of the stellar surface may be covered in spots without inducing any photometric modulation, if the spots are located in circumpolar regions that always face towards our line-of-sight, in longitudinally symmetric bands, or in many isotropic small spots.  

Coupled with the results of \citet{fang2016} and \citet{covey16}, the results shown here suggest that the contribution of starspots to pre-main sequence stellar evolution have been systematically underestimated.  
Some cool stars in the Pleiades have a large excess in Li abundance, some have inflated radii, some are fast rotators, and some have spots \citep[e.g.][]{somers15b}.  While the source overlap for these properties is unclear, these anomalies may all be explained if the pre-main sequence evolution of some stars is strongly affected by spots \citep{somers15}.  
LkCa 4 and other young stars with similar properties may help to connect the properties of anomalous Pleiades stars with earlier stages of stellar evolution.

The spectral inference technique described in this paper lays the foundation for future studies that seek accurate constraints on multiple photospheric components.  The near-IR has been shown to be the most sensitive region to search for cool emission from starspots.  Starspots and disk or accretion veiling can be modeled in this framework, and combining the two will enable studies of the large number of classical T-Tauri stars.  Applying this technique to large ensembles of T-Tauri stars across a wide range of mass and age will offer an observational picture of starspot evolution that can revise the current understanding of pre-main sequence stellar evolution.

\clearpage
\pagebreak

\appendix

\section{Further methodological details}
\label{methods-details}

When the synthetic stellar models used in \texttt{Starfish} are in absolute flux units, our mixture model is complete as stated in Equation \ref{eqn:mix_M}.  However \iancze\ employ \emph{standardized} $\flam$ when constructing the forward model $\vM$:
\begin{eqnarray} \label{eqn:normalization}
\bar \flam = \frac{\flam}{\int_{0}^{\infty} \flam d\lambda} = \frac{\flam}{f}
\end{eqnarray}

The reason for standardizing the flux is a matter of practicality: the number of PCA eigenspectra components in the spectral emulator scales steeply with the pixel-to-pixel variance of $f_{\lambda}(\{\vt_{\ast}\}^\textrm{grid})$, increasing the computational cost of spectral emulation.  The choice to standardize fluxes makes no difference for modeling a single photospheric component.  But for two-component photosphere models, the relative flux of the two model spectra needs to be accounted for to get an accurate estimate of the areal coverage fraction of the cool spots, $f_{\mathrm{cool}} \equiv \Omega_{\mathrm{cool}}/(\Omega_{\mathrm{hot}}+\Omega_{\mathrm{cool}})$.  So we scale the mixture model in the following way:

\begin{eqnarray} \label{eqn:norm_scaling}
f_{\lambda, \mathrm{mix}} &=& f_{\mathrm{hot}} \bar f_{\lambda, \mathrm{hot}} \times \Omega_{\mathrm{hot}} + f_{\mathrm{cool}} \bar f_{\lambda, \mathrm{cool}} \times \Omega_{\mathrm{cool}} \\
q &=& q(\vt_{\ast})\equiv \frac{f_{\mathrm{cool}}(\teffb, ...)}{f_{\mathrm{hot}}(\teffa, ...)} \\
f_{\lambda, \mathrm{mix}}^{\prime} &=& \bar f_{\lambda, \mathrm{hot}} \times \Omega_{\mathrm{hot}} + q \bar f_{\lambda, \mathrm{cool}} \times \Omega_{\mathrm{cool}}
\end{eqnarray}

where the prime symbol in the final line indicates a re-standardized mixture model flux, where the relative fluxes of the model components are now correctly scaled.

We are then tasked with computing an estimator, $\hat f(\vt_{\ast})$, for estimating the scale factor $q$ in-between model gridpoints.  One robust approach would be to follow \iancze\ by training a Gaussian process on the $f(\{\vt_{\ast}\}^\textrm{grid})$.  Instead, we simply linearly interpolated between the model grid-points.  Interpolation can cause pile-up near model grid-points, as noted in \citet{cottaar14}, which motivated the spectral emulation procedure in \iancze.  We assume the interpolation of $\hat f$ is smooth enough that we will not see such pileups and if even we did see pileups, they would mostly be discernable as kinks in the distribution of samples in the starspot coverage fraction $f_{\mathrm{cool}}$.  One drawback of our estimator method compared to the Gaussian Process regression method is that we do not propagate the uncertainty associated with the absolute flux ratio interpolation into our estimate of $f_{\mathrm{cool}}$.  We assume this uncertainty is relatively small and can be ignored.

The mixture model is a linear operation with all the same stellar extrinsic parameters, so we can re-use all the same post-processed eigenspectra $\widetilde{\mathbf{\Xi}}$, mean spectrum $\widetilde{\xi}_\mu$, and variance spectrum $\widetilde{\xi}_\sigma$, with the tildes representing all post processing \emph{except} the $\Omega$ scaling.  We calculate \emph{two} sets of eigenspectra weights $\mathbf{w}_{\in (\mathrm{hot}, \mathrm{cool})}$, and their associated mean and covariances following the Appendix of \iancze, and yielding:

\begin{equation}
  \mathsf{M}_{\mathrm{mix}}^\prime = \Omega_{\mathrm{hot}} (\widetilde{\xi}_\mu + \mathbf{X} \mathbf{\mu}_{\mathbf{w}, \mathrm{hot}}) + q \Omega_{\mathrm{cool}} (\widetilde{\xi}_\mu + \mathbf{X} \mathbf{\mu}_{\mathbf{w}, \mathrm{cool}})
\end{equation}

\begin{equation}
  \mathsf{C}_{\mathrm{mix}}^\prime = \Omega_{\mathrm{hot}}^2 \mathbf{X} \mathbf{\Sigma}_\mathbf{w, \mathrm{hot}} \mathbf{X}^T + q \Omega_{\mathrm{cool}}^2 \mathbf{X} \mathbf{\Sigma}_\mathbf{w, \mathrm{cool}} \mathbf{X}^T + \mathsf{C}
  \label{eqn:modC}
\end{equation}

\section{Further examples of full spectrum fitting}

\begin{figure*}
 \centering
 \includegraphics[width=0.95\textwidth]{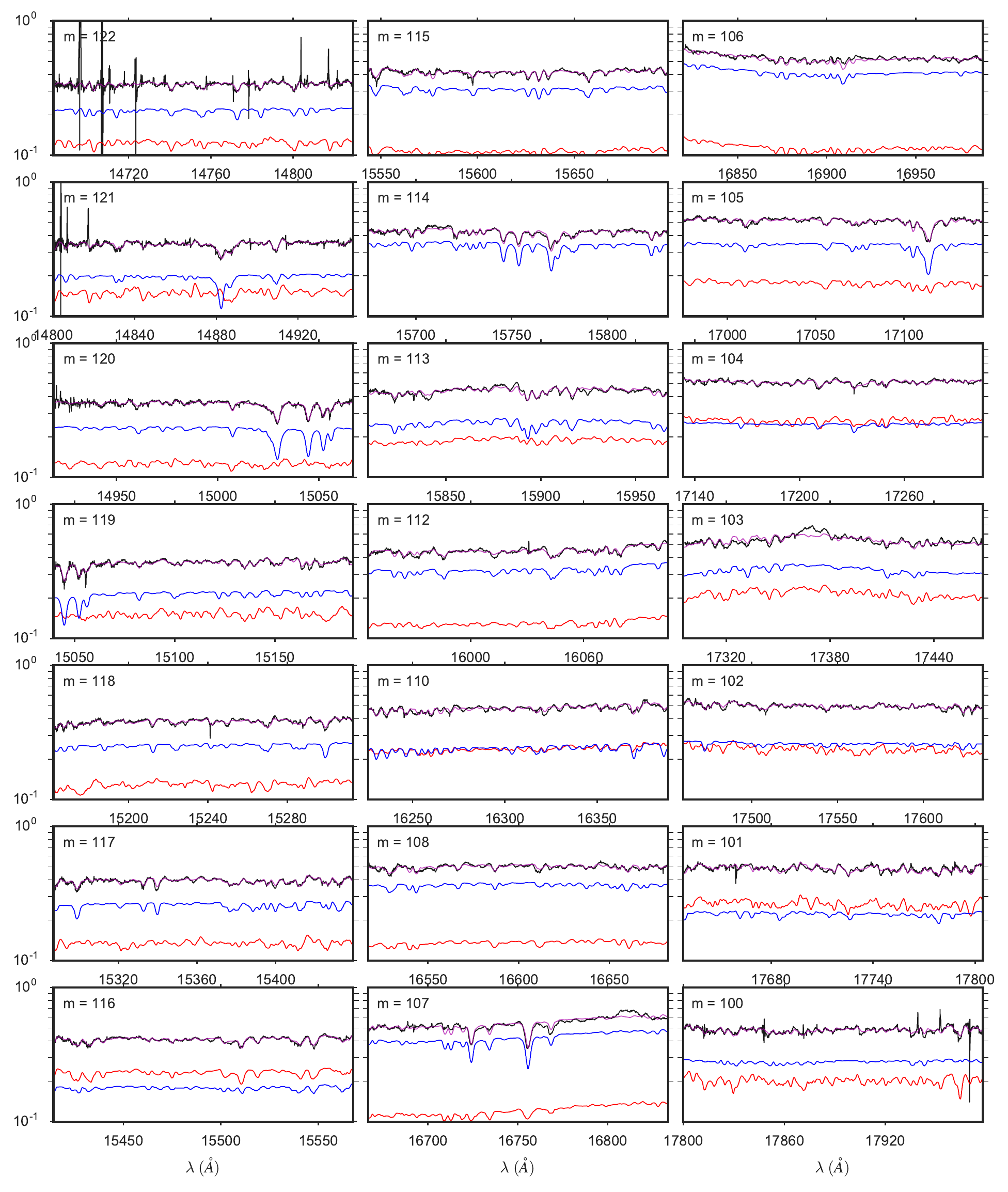}
 \caption{IGRINS Orders $122-100$, with panels arranged with the shortest wavelength in the upper left corner with central wavelength decreasing toward the bottom of the leftmost column, then decreasing through the subsequent columns.  The $y-$axis is on a logarithmic scale.  The red line is the cool photosphere while the blue line is the hot photosphere.  The purple line is the composite mixture model.}
 \label{fig:Hband3x7}
\end{figure*}

\begin{figure*}
 \centering
 \includegraphics[width=0.95\textwidth]{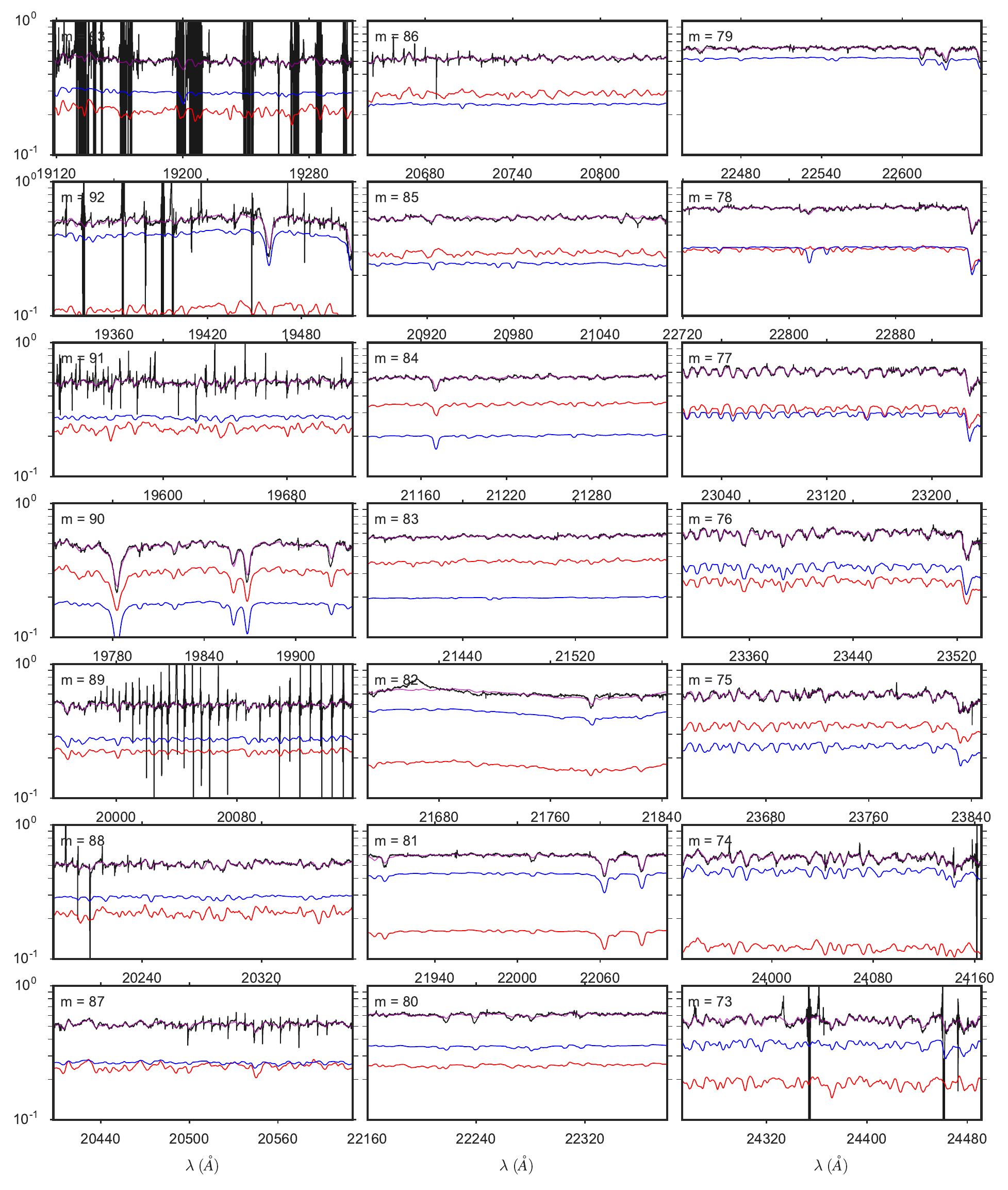}
 \caption{IGRINS Orders $93-73$, with the same layout, ordering, and colors as Figure \ref{fig:Hband3x7}. The $y-$axis is on a logarithmic scale.}
 \label{fig:Kband3x7}
\end{figure*}

Figures \ref{fig:Hband3x7} and \ref{fig:Kband3x7} show spectra from 42 IGRINS orders.

\section{Full table of IGRINS best fits}

Table \ref{tbl_order_results} lists all the IGRINS spectral order results.
\LongTables
%%%%%%%%%%%%%%%%%%%%%%%%%%%%%%%%%%%%%%%%
% TABLE - History of LkCa4
%%%%%%%%%%%%%%%%%%%%%%%%%%%%%%%%%%%%%%%%
\begin{deluxetable*}{ccccccc}

\tabcolsep=0.11cm
%\rotate
\tabletypesize{\footnotesize}
\tablecaption{Results from one- and two- component IGRINS spectral fitting\label{tbl_order_results}}
\tablewidth{0pt}
\tablehead{
& & & One comp. & \multicolumn{3}{c}{Two components} \\
\cline{5-7} 
\colhead{Order} &
\colhead{Instrument} &
\colhead{$\lambda_1-\lambda_2$} &
\colhead{$T_{\rm eff}$} &
\colhead{$T_{\rm hot}$} &
\colhead{$T_{\rm cool}$} &
\colhead{$f_{\rm cool}$} \\
\colhead{} &
\colhead{} &
\colhead{$\mu$m} &
\colhead{K} &
\colhead{K} &
\colhead{K} &
\colhead{}
}
\startdata
     0 &   ESPaDoNs &      5161$-$5200 &   $3734^{+183}_{-63}$ &               $\cdots$ &               $\cdots$ &                $\cdots$ \\
     1 &   ESPaDoNs &      5199$-$5263 &    $3991^{+70}_{-81}$ &               $\cdots$ &               $\cdots$ &                $\cdots$ \\
     2 &   ESPaDoNs &      5262$-$5328 &  $3990^{+196}_{-140}$ &               $\cdots$ &               $\cdots$ &                $\cdots$ \\
     3 &   ESPaDoNs &      5327$-$5395 &    $3923^{+72}_{-77}$ &               $\cdots$ &               $\cdots$ &                $\cdots$ \\
     4 &   ESPaDoNs &      5394$-$5465 &    $3812^{+54}_{-85}$ &               $\cdots$ &               $\cdots$ &                $\cdots$ \\
     5 &   ESPaDoNs &      5464$-$5537 &    $4165^{+59}_{-61}$ &               $\cdots$ &               $\cdots$ &                $\cdots$ \\
     6 &   ESPaDoNs &      5536$-$5612 &    $4129^{+37}_{-38}$ &               $\cdots$ &               $\cdots$ &                $\cdots$ \\
     7 &   ESPaDoNs &      5611$-$5689 &    $4057^{+64}_{-52}$ &               $\cdots$ &               $\cdots$ &                $\cdots$ \\
     8 &   ESPaDoNs &      5688$-$5770 &    $4038^{+50}_{-44}$ &               $\cdots$ &               $\cdots$ &                $\cdots$ \\
     9 &   ESPaDoNs &      5769$-$5854 &    $4212^{+71}_{-64}$ &               $\cdots$ &               $\cdots$ &                $\cdots$ \\
    10 &   ESPaDoNs &      5853$-$5941 &    $4119^{+52}_{-46}$ &               $\cdots$ &               $\cdots$ &                $\cdots$ \\
    11 &   ESPaDoNs &      5940$-$6032 &    $4091^{+83}_{-63}$ &               $\cdots$ &               $\cdots$ &                $\cdots$ \\
    12 &   ESPaDoNs &      6031$-$6127 &    $4108^{+46}_{-42}$ &               $\cdots$ &               $\cdots$ &                $\cdots$ \\
    13 &   ESPaDoNs &      6126$-$6225 &    $3941^{+28}_{-24}$ &               $\cdots$ &               $\cdots$ &                $\cdots$ \\
    14 &   ESPaDoNs &      6225$-$6329 &    $4012^{+33}_{-42}$ &               $\cdots$ &               $\cdots$ &                $\cdots$ \\
    15 &   ESPaDoNs &      6328$-$6436 &    $4238^{+49}_{-40}$ &               $\cdots$ &               $\cdots$ &                $\cdots$ \\
    16 &   ESPaDoNs &      6435$-$6549 &    $3880^{+61}_{-76}$ &               $\cdots$ &               $\cdots$ &                $\cdots$ \\
    17 &   ESPaDoNs &      6548$-$6667 &    $3874^{+72}_{-70}$ &               $\cdots$ &               $\cdots$ &                $\cdots$ \\
    18 &   ESPaDoNs &      6666$-$6791 &    $4054^{+42}_{-32}$ &               $\cdots$ &               $\cdots$ &                $\cdots$ \\
    20 &   ESPaDoNs &      6920$-$7057 &    $3907^{+32}_{-20}$ &               $\cdots$ &               $\cdots$ &                $\cdots$ \\
    21 &   ESPaDoNs &      7056$-$7201 &    $3910^{+25}_{-20}$ &               $\cdots$ &               $\cdots$ &                $\cdots$ \\
    22 &   ESPaDoNs &      7200$-$7352 &    $3878^{+28}_{-15}$ &               $\cdots$ &               $\cdots$ &                $\cdots$ \\
    23 &   ESPaDoNs &      7351$-$7512 &    $4085^{+51}_{-66}$ &               $\cdots$ &               $\cdots$ &                $\cdots$ \\
    26 &   ESPaDoNs &      7858$-$8051 &    $3852^{+24}_{-24}$ &               $\cdots$ &               $\cdots$ &                $\cdots$ \\
    29 &   ESPaDoNs &      8473$-$8707 &    $3504^{+49}_{-46}$ &               $\cdots$ &               $\cdots$ &                $\cdots$ \\
    30 &   ESPaDoNs &      8706$-$8954 &    $3821^{+33}_{-29}$ &               $\cdots$ &               $\cdots$ &                $\cdots$ \\
   123 &     IGRINS &    14519$-$14721 &              $\cdots$ &    $4182^{+107}_{-75}$ &    $2773^{+254}_{-68}$ &  $0.82^{+0.03}_{-0.72}$ \\
   122 &     IGRINS &    14634$-$14837 &              $\cdots$ &   $4185^{+187}_{-137}$ &   $2819^{+209}_{-105}$ &  $0.78^{+0.04}_{-0.22}$ \\
   121 &     IGRINS &    14750$-$14956 &              $\cdots$ &   $4084^{+153}_{-144}$ &     $2762^{+83}_{-58}$ &  $0.79^{+0.04}_{-0.05}$ \\
   120 &     IGRINS &    14868$-$15076 &              $\cdots$ &    $4027^{+76}_{-112}$ &     $2739^{+94}_{-37}$ &  $0.69^{+0.05}_{-0.08}$ \\
   119 &     IGRINS &    14988$-$15198 &              $\cdots$ &   $4093^{+123}_{-132}$ &     $2719^{+67}_{-18}$ &  $0.76^{+0.04}_{-0.05}$ \\
   118 &     IGRINS &    15111$-$15322 &              $\cdots$ &   $4124^{+142}_{-131}$ &     $2730^{+83}_{-30}$ &  $0.74^{+0.08}_{-0.08}$ \\
   117 &     IGRINS &    15235$-$15448 &  $4034^{+147}_{-122}$ &     $4105^{+63}_{-73}$ &     $2731^{+70}_{-29}$ &  $0.73^{+0.05}_{-0.06}$ \\
   116 &     IGRINS &    15362$-$15577 &    $4239^{+55}_{-73}$ &   $4121^{+147}_{-146}$ &    $2795^{+169}_{-85}$ &  $0.80^{+0.04}_{-0.09}$ \\
   115 &     IGRINS &    15490$-$15707 &   $3718^{+94}_{-130}$ &   $3914^{+176}_{-151}$ &    $2769^{+298}_{-64}$ &  $0.55^{+0.14}_{-0.44}$ \\
   114 &     IGRINS &    15622$-$15841 &              $\cdots$ &   $4021^{+159}_{-122}$ &    $2751^{+147}_{-43}$ &  $0.64^{+0.10}_{-0.13}$ \\
   113 &     IGRINS &    15755$-$15976 &              $\cdots$ &   $3874^{+265}_{-170}$ &    $2779^{+193}_{-75}$ &  $0.77^{+0.08}_{-0.11}$ \\
   112 &     IGRINS &    15891$-$16114 &              $\cdots$ &   $3986^{+268}_{-206}$ &   $2943^{+425}_{-198}$ &  $0.53^{+0.16}_{-0.17}$ \\
   111 &     IGRINS &    16030$-$16255 &  $4101^{+100}_{-116}$ &               $\cdots$ &               $\cdots$ &                $\cdots$ \\
   110 &     IGRINS &    16171$-$16398 &   $4233^{+59}_{-104}$ &   $4180^{+264}_{-193}$ &    $2777^{+184}_{-68}$ &  $0.83^{+0.06}_{-0.17}$ \\
   109 &     IGRINS &    16314$-$16543 &   $3933^{+126}_{-97}$ &          $0^{+0}_{-0}$ &          $0^{+0}_{-0}$ &  $0.50^{+0.00}_{-0.00}$ \\
   108 &     IGRINS &    16461$-$16692 &  $4150^{+100}_{-176}$ &   $4248^{+183}_{-193}$ &    $2802^{+339}_{-84}$ &  $0.71^{+0.12}_{-0.35}$ \\
   107 &     IGRINS &    16610$-$16843 &  $3750^{+186}_{-112}$ &   $3813^{+521}_{-177}$ &   $3237^{+437}_{-486}$ &  $0.38^{+0.27}_{-0.36}$ \\
   106 &     IGRINS &    16762$-$16997 &    $4225^{+68}_{-92}$ &   $4324^{+161}_{-168}$ &  $2882^{+1227}_{-175}$ &  $0.53^{+0.19}_{-0.53}$ \\
   105 &     IGRINS &    16917$-$17155 &               $<3711$ &   $3930^{+209}_{-152}$ &    $2763^{+212}_{-59}$ &  $0.66^{+0.06}_{-0.10}$ \\
   104 &     IGRINS &    17075$-$17315 &               $<3776$ &   $4300^{+177}_{-245}$ &   $2870^{+115}_{-130}$ &  $0.83^{+0.05}_{-0.05}$ \\
   103 &     IGRINS &    17236$-$17478 &               $<3700$ &   $3848^{+312}_{-334}$ &   $2972^{+199}_{-234}$ &  $0.65^{+0.18}_{-0.30}$ \\
   102 &     IGRINS &    17400$-$17645 &               $<3629$ &   $4336^{+136}_{-159}$ &    $2751^{+113}_{-48}$ &  $0.82^{+0.04}_{-0.05}$ \\
   101 &     IGRINS &    17568$-$17815 &               $<3543$ &   $4199^{+264}_{-574}$ &   $2916^{+158}_{-158}$ &  $0.79^{+0.05}_{-0.23}$ \\
   100 &     IGRINS &    17739$-$17988 &               $<3598$ &   $4312^{+161}_{-193}$ &    $2765^{+135}_{-61}$ &  $0.79^{+0.04}_{-0.06}$ \\
    99 &     IGRINS &    17914$-$18165 &               $<3552$ &               $\cdots$ &               $\cdots$ &                $\cdots$ \\
    94 &     IGRINS &    18855$-$19117 &              $\cdots$ &  $4293^{+185}_{-1132}$ &   $2899^{+304}_{-150}$ &  $0.81^{+0.07}_{-0.39}$ \\
    93 &     IGRINS &    19053$-$19318 &              $\cdots$ &   $4044^{+353}_{-378}$ &    $2740^{+119}_{-36}$ &  $0.81^{+0.08}_{-0.19}$ \\
    92 &     IGRINS &    19256$-$19524 &              $\cdots$ &   $3770^{+155}_{-183}$ &   $2859^{+510}_{-149}$ &  $0.56^{+0.18}_{-0.21}$ \\
    91 &     IGRINS &    19462$-$19734 &    $3464^{+70}_{-67}$ &   $3962^{+389}_{-426}$ &   $3126^{+857}_{-325}$ &  $0.68^{+0.18}_{-0.56}$ \\
    90 &     IGRINS &    19674$-$19948 &    $3594^{+75}_{-70}$ &   $3907^{+251}_{-211}$ &    $3463^{+90}_{-161}$ &  $0.75^{+0.10}_{-0.28}$ \\
    89 &     IGRINS &    19890$-$20168 &    $3557^{+82}_{-72}$ &   $3706^{+254}_{-151}$ &   $3412^{+183}_{-617}$ &  $0.45^{+0.20}_{-0.23}$ \\
    88 &     IGRINS &    20112$-$20392 &    $3500^{+61}_{-60}$ &   $3917^{+208}_{-129}$ &    $2812^{+250}_{-99}$ &  $0.75^{+0.09}_{-0.06}$ \\
    87 &     IGRINS &    20338$-$20622 &    $3314^{+77}_{-90}$ &   $3709^{+305}_{-207}$ &   $2848^{+167}_{-119}$ &  $0.75^{+0.09}_{-0.24}$ \\
    86 &     IGRINS &    20570$-$20857 &    $3544^{+51}_{-48}$ &   $3915^{+475}_{-275}$ &   $3135^{+281}_{-345}$ &  $0.68^{+0.15}_{-0.18}$ \\
    85 &     IGRINS &    20808$-$21098 &   $3315^{+99}_{-107}$ &   $4220^{+252}_{-322}$ &   $2828^{+205}_{-112}$ &  $0.85^{+0.05}_{-0.10}$ \\
    84 &     IGRINS &    21051$-$21345 &    $3475^{+62}_{-56}$ &   $3681^{+444}_{-191}$ &   $3414^{+125}_{-130}$ &  $0.68^{+0.17}_{-0.12}$ \\
    83 &     IGRINS &    21300$-$21597 &   $3234^{+81}_{-108}$ &   $4334^{+144}_{-235}$ &   $2896^{+166}_{-137}$ &  $0.90^{+0.03}_{-0.03}$ \\
    82 &     IGRINS &    21555$-$21856 &      $3589^{+2}_{-8}$ &   $3535^{+582}_{-258}$ &   $2964^{+297}_{-241}$ &  $0.62^{+0.21}_{-0.38}$ \\
    81 &     IGRINS &    21817$-$22121 &   $3466^{+117}_{-87}$ &   $3536^{+146}_{-124}$ &   $3373^{+152}_{-150}$ &  $0.34^{+0.16}_{-0.10}$ \\
    80 &     IGRINS &    22085$-$22393 &  $3157^{+132}_{-124}$ &   $3543^{+404}_{-273}$ &    $2797^{+336}_{-88}$ &  $0.69^{+0.16}_{-0.21}$ \\
    79 &     IGRINS &    22360$-$22671 &    $3437^{+79}_{-60}$ &    $3511^{+138}_{-92}$ &   $3383^{+122}_{-120}$ &  $0.30^{+0.13}_{-0.10}$ \\
    78 &     IGRINS &    22643$-$22957 &    $3197^{+43}_{-43}$ &   $3808^{+321}_{-231}$ &    $2727^{+159}_{-26}$ &  $0.78^{+0.07}_{-0.06}$ \\
    77 &     IGRINS &    22932$-$23251 &    $3273^{+52}_{-49}$ &   $4221^{+254}_{-392}$ &    $2716^{+273}_{-15}$ &  $0.85^{+0.03}_{-0.12}$ \\
    76 &     IGRINS &    23230$-$23552 &    $3321^{+54}_{-53}$ &   $3428^{+317}_{-116}$ &   $3123^{+193}_{-362}$ &  $0.46^{+0.22}_{-0.21}$ \\
    75 &     IGRINS &    23535$-$23861 &              $\cdots$ &   $3649^{+528}_{-218}$ &   $2956^{+394}_{-222}$ &  $0.62^{+0.17}_{-0.17}$ \\
    74 &     IGRINS &    23849$-$24179 &              $\cdots$ &    $3523^{+133}_{-94}$ &   $3053^{+388}_{-308}$ &  $0.34^{+0.20}_{-0.19}$ \\
    73 &     IGRINS &    24172$-$24505 &              $\cdots$ &   $4119^{+345}_{-433}$ &   $3029^{+244}_{-294}$ &  $0.80^{+0.08}_{-0.13}$ \\
    72 &     IGRINS &    24503$-$24840 &              $\cdots$ &   $3897^{+398}_{-184}$ &   $2881^{+425}_{-157}$ &  $0.74^{+0.11}_{-0.26}$ \\
\enddata
\tablecomments{The uncertainties are the 5$^{th}$ and 95$^{th}$ percentiles of the marginalized samples, corresponding roughly to ``2 $\sigma$'' error bars.}
%\tablerefs{}

%\end{deluxetable*}
\end{deluxetable*}

\section{Previous work}

Table \ref{tbl_history} lists measurements of \name from previous studies, compared to this work.
%%%%%%%%%%%%%%%%%%%%%%%%%%%%%%%%%%%%%%%%
% TABLE - History of LkCa4
%%%%%%%%%%%%%%%%%%%%%%%%%%%%%%%%%%%%%%%%
%\begin{deluxetable*}{lccccccccc}
\begin{deluxetable}{p{4cm}ccccccccc}

\tabcolsep=0.11cm
%\rotate
\tabletypesize{\footnotesize}
\tablecaption{Previous studies of \name \label{tbl_history}}
\tablewidth{0pt}
\tablehead{
\colhead{Reference} &
\colhead{Band(s)} &
\colhead{R} &
\colhead{SpT} &
\colhead{$T_{\rm eff}$} &
\colhead{$\log{g}$} &
\colhead{$A_V$} &
\colhead{$v\sin{i}$} &
\colhead{$v_{z}$} &
\colhead{$\log{L/L_{\odot}}$}
\\
\colhead{} &
\colhead{} &
\colhead{$\lambda/\delta \lambda$} &
\colhead{} &
\colhead{K} &
\colhead{} &
\colhead{} &
\colhead{km/s} &
\colhead{km/s} &
\colhead{}
}
\startdata
 \citet{herbig86} & $V$ & 100000 & K7 V & - & - & 0.68 & 26.1$\pm$2.4 & 13$\pm$4 & 0.02\\
 \citet{hartmann87} & $U$ & 100000 & - & - & - & - & 26.1$\pm$2.4 & 16.9$\pm$2.6 & \\
 \citet{downes88} & $V$ & 13$\AA$ & Me & - & - &  - & - & - & \\
 \citet{strom89a} & IRAS & - & K7:V & - & - & 0.95 & - & - & 0.04\\
 \citet{strom89b} & $V$ & 3500 & K7:V & - & - & - & - & - & \\
 \citet{stauffer91} & $VRI$ & 500 & M1 & - & - & - & - &  - & \\
 \citet{strom94} & $V$ & - & K7 & 4000 & - & 1.25 & - & - &  0.06\\
 \citet{kenyon95} & $U$-IRAS & - & K7 & 4060 & - & 0.69 & - & - & -0.08\\
 \citet{hartigan95} & V & 25000 & K7 & 4000 & - & 0.68  & - & - & -0.07\\
 \citet{white01} & $VI$ & - & K7 & - & - & 1.21 & - & -  & -0.02 \\
 \citet{nguyen09} & $BVRI$ & 60000 & - & - & - & - & $30\pm2$ & - & \\
 \citet{nguyen12} & $BVRI$ & 60000 & - & - & - & - & $30\pm2$ & $16.0\pm4.0$ & \\
 \citet{grankin13} & $BVRI$ & - & K7 & 4040 & - & 0.54 & 26.1 & - & -0.13 \\
 \citet{donati14} & $UBVRI$ & 68000 & - & 4100$\pm50$ & 3.8$\pm$0.1 & 0.68$\pm$0.15 & 28.0$\pm$0.5 & 16.8$\pm$0.1 & -0.04$\pm$0.11 \\
 \citet{herczeg14} & $UBVRI$ & 700 & M1.3 & 3670 & - & 0.35 & - & - & -0.29 \\
 \emph{This Work: Best} & $HK$ & 45000 & - & 3180 & $\sim$3.8 & 0.3 & 29$\pm$3 & - &  -0.26 \\
 \emph{This Work: Alternate} & $HK$ & 45000 & - & 3330 & $\sim$3.8 & - & 29$\pm$3 & - & $\sim-0.26$ \\
 
\enddata

%\end{deluxetable*}
\end{deluxetable}

\acknowledgements

We thank the anonymous referee, Chris Kochanek, and Daniel T. Jaffe for valuable comments on the draft.  MG-S and GJH also thank Steve Saar, Lisa Prato, and Lynne Hillenbrand for valuable discussions about spots.

Portions of the data, Python code, Jupyter notebooks, Starfish configuration files, MCMC samples, and revision history of this manuscript are available on the project's GitHub repository at \url{https://github.com/BrownDwarf/welter}.

MG-S and GJH are supported by general grant 11473005 awarded by the National Science Foundation of China.   The ESPaDOnS observations are supported by the contribution to the MaTYSSE Large Project on CFHT obtained through the Telescope Access Program (TAP), which has been funded by the ``the Strategic Priority Research Program---The Emergence of Cosmological Structures'' of the Chinese Academy of Sciences (Grant No.11 XDB09000000) and the Special Fund for Astronomy from the Ministry of Finance. 
TW-SH is supported by the DOE Computational Science Graduate Fellowship, grant number DE-FG02-97ER25308.

We thank LCOGT and its staff for their continued support of ASAS-SN. ASAS-SN is supported by NSF grant AST-1515927. Development of ASAS-SN has been supported by NSF grant AST-0908816, the Center for Cosmology and AstroParticle Physics at the Ohio State University, the Mt. Cuba Astronomical Foundation, and by George Skestos. 

BJS is supported by NASA through Hubble Fellowship grant HST-HF-51348.001 awarded by the Space Telescope Science Institute, which is operated by the Association of Universities for Research in Astronomy, Inc., for NASA, under contract NAS 5-26555.  

This work used the Immersion Grating Infrared Spectrograph (IGRINS) that was developed under a collaboration between the University of Texas at Austin and the Korea Astronomy and Space Science Institute (KASI) with the financial support of the US National Science Foundation under grant ASTR1229522, of the University of Texas at Austin, and of the Korean GMT Project of KASI.

This research has made use of NASA's Astrophysics Data System.  The research is based on data from the OMC Archive at CAB (INTA-CSIC), pre-processed by ISDC.
We acknowledge with thanks the variable star observations from the AAVSO International Database contributed by observers worldwide and used in this research.
This publication makes use of data products from the Two Micron All Sky Survey, which is a joint project of the University of Massachusetts and the Infrared Processing and Analysis Center/California Institute of Technology, funded by the National Aeronautics and Space Administration and the National Science Foundation.

{\it Facilities:} \facility{Smith (IGRINS)}, \facility{AAVSO}, \facility{CFHT (ESPaDOnS)}, \facility{INTEGRAL (OMC)}, \facility{ASAS}, \facility{CrAO:1.25m}, \facility{ARC (TripleSpec)}, \facility{Hale (DBSP)}, \facility{Gaia}

{\it Software: } 
 \project{pandas} \citep{mckinney10},
 \project{emcee} \citep{foreman13},
 \project{matplotlib} \citep{hunter07},
 \project{numpy} \citep{vanderwalt11},
 \project{scipy} \citep{jones01},
 \project{ipython} \citep{perez07},
 \project{gatspy} \citep{JakeVanderplas2015},
 \project{starfish} \citep{czekala15},
 \project{seaborn} \citep{waskom14}
%\software{%
% \project{pandas} \citep{mckinney10}
%    \project{emcee} \citep{foreman13},
% \project{matplotlib} \citep{hunter07},
% \project{numpy} \citep{vanderwalt11},
% \project{scipy} \citep{jones01},
% \project{ipython} \citep{perez07},
% \project{gatspy} \citep{JakeVanderplas2015},
% \project{starfish} \citep{czekala15}}.

\clearpage

\bibliographystyle{apj}
\bibliography{ms}

\end{document}